\documentclass[final,3p,11pt,pdflatex]{elsarticle}
\usepackage[utf8x]{inputenc}      
\usepackage{amsmath,amssymb}
\usepackage[T1]{fontenc}          
\usepackage{booktabs,tabularx}
\usepackage{rotating}             
\usepackage{xspace}
\usepackage[usenames]{xcolor}
\usepackage{tikz,tikz-uml}
\usepackage{listings}
\usepackage[absolute]{textpos}
\bibstyle{elsarticle-num}
\usepackage[pdftitle={FlexibleSUSY --- A spectrum generator generator for supersymmetric models},
pdfauthor={Peter Athron,Jae-hyeon Park,Dominik Stockinger,Alexander Voigt},
pdfkeywords={FlexibleSUSY,supersymmetry,spectrum,generator,MSSM,NMSSM,E6SSM},
bookmarks=true, linktocpage]{hyperref}

\newcommand{\sarah}{SARAH\@\xspace}
\newcommand{\fs}{FlexibleSUSY\@\xspace}
\newcommand{\mathematica}{Mathematica\xspace}
\newcommand{\ESSM}{E$_6$SSM\@\xspace}
\newcommand{\code}[1]{\lstinline|#1|}  
\newcommand{\textoverline}[1]{$\overline{\mbox{#1}}$}
\newcommand{\DRbar}{\textoverline{DR}\xspace}
\newcommand{\MSbar}{\textoverline{MS}\xspace}
\newcommand{\unit}[1]{\,\text{#1}}      
\newcommand{\userinput}{\text{input}}
\newcommand{\pole}{\text{pole}}
\newcommand{\Lagr}{\mathcal{L}}
\newcommand{\unity}{\mathbf{1}}
\newcommand{\figref}[1]{\figurename~\ref{#1}}
\newcommand{\secref}[1]{Section~\ref{#1}}
\newcommand{\tabref}[1]{\tablename~\ref{#1}}
\newcommand{\ptitle}[1]{\emph{#1}}
\renewcommand{\ptitle}[1]{}
\newcommand{\Zv}{\mathbf{\backslash}\mkern-11.0mu{Z}}
\makeatletter
\lstnewenvironment{numlstlisting}[1][]{%
  \lstset{%
    #1,
    numbers=left,
    firstnumber=auto,
    numberstyle=\tiny\sffamily}%
  \csname\@lst @SetFirstNumber\endcsname
}{%
  \csname \@lst @SaveFirstNumber\endcsname
}
\makeatother
\DeclareMathOperator{\diag}{diag}
\DeclareMathOperator{\sign}{sign}
\DeclareMathOperator{\re}{Re}

\def\at{\alpha_t}
\def\ab{\alpha_b}
\def\as{\alpha_s}
\def\atau{\alpha_{\tau}}

\def\oatab{O(\at\ab)}
\def\oatas{O(\at\as)}
\def\oabas{O(\ab\as)}

\def\oatq{O(\at^2)}
\def\oabq{O(\ab^2)}
\def\oatauq{O(\atau^2)}

\def\oatplusabsq{O((\at+\ab)^2)}

\journal{Computer Physics Communications}
\begin{document}
\begin{frontmatter}
 \title{\Large\bf FlexibleSUSY --- A spectrum generator generator for supersymmetric models}

\author[adelaide]{Peter Athron}
\author[valencia]{Jae-hyeon Park}
\author[dresden]{Dominik St\"ockinger}
\author[desy]{Alexander Voigt\corref{cor1}}
\ead{Alexander.Voigt@desy.de}
\cortext[cor1]{Corresponding author}
\address[adelaide]{ARC Centre of Excellence for Particle Physics at 
the Tera-scale, School of Chemistry and Physics, University of Adelaide, 
Adelaide SA 5005 Australia}
\address[valencia]{Departament de F\'{i}sica Te\`{o}rica and IFIC,
Universitat de Val\`{e}ncia-CSIC,
46100, Burjassot, Spain}
\address[dresden]{Institut f\"ur Kern- und Teilchenphysik,
TU Dresden, Zellescher Weg 19, 01069 Dresden, Germany}
\address[desy]{Deutsches Elektronen-Synchrotron (DESY), 22607 Hamburg, Germany}
   
  \begin{abstract}
    We introduce \fs, a \mathematica and C++ package, which generates a fast,
    precise C++ spectrum generator for any SUSY model specified by the
    user.  The generated code is designed with both speed and
    modularity in mind, making it easy to adapt and extend with new
    features. The model is specified by supplying the superpotential,
    gauge structure and particle content in a \sarah model file;
    specific boundary conditions e.g.\ at the GUT, weak or
    intermediate scales are defined in a separate \fs model file.
    From these model files, \fs generates C++ code for self-energies,
    tadpole corrections, renormalization group equations (RGEs) and
    electroweak symmetry breaking (EWSB) conditions and combines them
    with numerical routines for solving the RGEs and EWSB conditions
    simultaneously.  The resulting spectrum generator is then able to
    solve for the spectrum of the model, including loop-corrected pole
    masses, consistent with user specified boundary conditions.  The
    modular structure of the generated code allows for individual
    components to be replaced with an alternative if available. \fs
    has been carefully designed to grow as alternative solvers and
    calculators are added.  Predefined models include the MSSM, NMSSM,
    \ESSM, USSM, $R$-symmetric models and models with right-handed
    neutrinos.
  \end{abstract}

\end{frontmatter}

\begin{textblock*}{7em}(\textwidth,1cm)
\noindent\footnotesize
FTUV--14--3904 \\
IFIC--14--40 \\
ADP-14-17/T875
\end{textblock*}

\clearpage
\section{Program Summary}
\noindent{\em Program title:} \fs\\[0.5em] {\em Program obtainable from:}
         {\tt http://flexiblesusy.hepforge.org/}\\[0.5em] {\em Distribution
           format:}\/ tar.gz\\[0.5em] {\em Programming language:} {\tt
           C++, Wolfram/\mathematica, FORTRAN, Bourne shell}\\[0.5em]
         {\em Computer:}\/ Personal computer\\[0.5em] {\em Operating
           system:}\/ Tested on Linux 3.x, Mac OS X\\[0.5em]
         {\em External routines:}\/ SARAH 4.0.4, Boost library,
         Eigen, LAPACK\\[0.5em] {\em
           Typical running time:}\/ 0.06--0.2 seconds per parameter
         point\\[0.5em] {\em Nature of problem:}\/ Determining the mass
         spectrum and mixings for any supersymmetric model. The
         generated code must find simultaneous solutions to
         constraints which are specified at two or more different
         renormalization scales, which are connected by
         renormalization group equations forming a large set of
         coupled first-order differential equations. \\[0.5em] {\em Solution method:}\/
         Nested iterative algorithm and numerical minimization of the
         Higgs potential.  \\[0.5em] {\em Restrictions:} The couplings must
         remain perturbative at all scales between the highest and
         lowest boundary condition.  \fs~ assumes that all couplings
         of the model are real (i.e.\ $CP$-conserving). Due to the
         modular nature of the generated code adaption and extension
         to overcome restrictions in scope is quite straightforward.

\clearpage
\tableofcontents

\newpage
\section{Introduction}

Supersymmetry (SUSY) provides the only non-trivial way to extend the
space-time symmetries of the Poincar\'e
group, which still has scattering in the resulting quantum theory \cite{Coleman:1967ad,Haag:1974qh}. This leads many to suspect that
SUSY may be realized in nature in some form. In particular
supersymmetric extensions of the Standard Model (SM) where SUSY is broken
at the TeV scale have been proposed to solve the hierarchy
problem \cite{Weinberg:1975gm, Weinberg:1979bn, Gildener:1976ai,
  Susskind:1978ms, 'tHooft:1980xb}, allow gauge coupling
unification \cite{Langacker:1990jh, Ellis:1990wk, Amaldi:1991cn,
  Langacker:1991an, Giunti:1991ta} and predict a dark matter candidate
which can fit the observed relic
density \cite{Goldberg:1983nd,Ellis:1983ew}.  Such models have also
been used for baryogenesis or leptogensis to solve the
matter-anti-matter asymmetry of the universe and have been considered
as the low energy effective models originating from string
theory.

Detailed phenomenological studies have been carried out for scenarios
within the minimal supersymmetric Standard Model (MSSM), for a review
see \cite{Chung:2003fi}.  Such work has been greatly aided by public
spectrum generators for the MSSM
\cite{Allanach:2001kg,Porod:2003um,Djouadi:2002ze,Baer:1993ae,Chowdhury:2011zr},
allowing fast and reliable exploration of the sparticle spectrum,
mixings and couplings, which can be obtained from particular choices
of breaking mechanism inspired boundary conditions and specified
parameters. Beyond the MSSM there are also two public spectrum
generators \cite{Ellwanger:2004xm, Ellwanger:2005dv,
  Ellwanger:2006rn,Ellwanger:2008py, Allanach:2013kza} for the next to
minimal supersymmetric Standard Model (NMSSM)  (for
recent reviews see \cite{Ellwanger:2009dp, Maniatis:2009re}).

None of the fundamental motivations of supersymmetry requires
minimality, and specific alternatives to (or extensions of) the MSSM
are, for example, motivated by the $\mu$-problem of the MSSM
\cite{Kim:1983dt}; explaining the family structure (see
e.g.~\cite{King:2014nza}) or for successful baryogenesis or
leptogenesis (see e.g.~\cite{King:2008qb}). However constructing
specialized tools to study all relevant models would require an
enormous amount of work.  So general tools which can automate this
process and produce fast and reliable programs can greatly enhance our
ability to understand and test non-minimal realizations of
supersymmetry.

Recent experimental developments have also increased the relevancy of
such a tool. From the recent $7$ TeV and $8$ TeV runs at the Large
Hadron Collider (LHC) there have been two important developments.
Firstly low energy signatures expected from such models, such as the
classic jets plus missing transverse energy signature, have not been
observed, substantially raising the lower limit on sparticle masses
(see e.g.~\cite{Aad:2013wta,Chatrchyan:2014lfa}). No other signature
of beyond the Standard Model (BSM) physics has been observed, leaving
the fundamental questions which motivated BSM physics
unanswered. Secondly ATLAS and CMS discovered \cite{ATLAS:2012ae,
  Chatrchyan:2012tx} a light Higgs of $125$ GeV, within the mass range
that could be accommodated in the MSSM but requiring stops which are
significantly heavier than both the direct collider limits and
indirect limits that appears in constrained models from the
significantly higher limits on first and second generation squarks.

These developments motivate the exploration of non-minimal SUSY models
which ameliorate the naturalness problems by raising the tree level
Higgs mass, as can happen in the USSM
\cite{Fayet:1977yc,Cvetic:1997ky,Langacker:2008yv}, \ESSM
\cite{King:2005jy,Athron:2010zz} and similar models
\cite{Nevzorov:2012hs}, or from other gauge extensions
\cite{Batra:2003nj, Bharucha:2013ela}. At the same time they can also
motivate models that are developed with a fresh perspective, based on
other considerations.  In both cases exploration of such models can be
aided if it is possible to quickly create a fast spectrum generator.
Currently there is only one option for this, a SPheno-like FORTRAN
code which can be generated from \sarah
\cite{Staub:2010ty,Staub:2009bi,Staub:2010jh,Staub:2012pb,Staub:2013tta}.

\fs provides a much needed alternative to this with a structure which
has been freshly designed to accommodate as general range of models as
possible and to be easily adaptable to changing goals and new
ideas. \fs is a \mathematica and C++ package which uses \sarah to create a
fast, modular C++ spectrum generator for a user specified SUSY model.
The generated code structure is designed to be as flexible as possible
to accommodate different types of extensions and due to its modular
nature it is easy to modify, add new features and combine with other
programs.  The generated code has been extensively tested against well
known spectrum generators. As well as providing a solution for new
SUSY models, the generated MSSM and NMSSM codes offer a modern and fast
alternative to the existing public spectrum generators.

In \secref{sec:Program} we describe the program in more detail and
explain our design goals.  In \secref{sec:download} information on how
to download and compile the code may be found along with details on
how to get started quickly.  In \secref{sec:modfile} we describe how
the user can create a new \fs model file. A detailed
description of the structure and features of the generated code is
then given in \secref{sec:SpecGenStruct}.  In \secref{sec:Flexible} we
describe the various ways the code can be modified both at the meta
code level by writing model files and at the C++ code level by
modifying the code or adding new modules. Finally in
\secref{sec:comparison} we describe detailed comparisons between our
generated code and existing public spectrum generators as well as
against the SPheno-like FORTRAN code which can be created using SARAH.

\section{Overview of the program and design goals}
\label{sec:Program}

To study the properties of SUSY models programs are needed which
numerically calculate the pole masses and couplings of the SUSY
particles given a set of theory input parameters.  The output of these
so-called spectrum generators can be transferred to programs which
calculate further observables such as branching ratios or the dark
matter relic density.

In order to create a spectrum generator the SUSY model must be defined
by specifying the gauge group, the field content and mixings as well
as the superpotential and the soft-breaking terms.  From this
information the renormalization group equations, mass matrices,
self-energies, tadpole diagrams and electroweak symmetry breaking
(EWSB) conditions have to be derived.  These expressions must then be
combined in a computer program to allow for a numeric calculation of
the mass spectrum.  In addition most SUSY models require boundary
conditions for the model parameters at a low and a high scale.  For
example in the CMSSM mSUGRA boundary conditions for the soft-breaking
parameters are imposed at the gauge coupling unification scale.
Furthermore, at the $Z$ mass scale the CMSSM is matched to the Standard
Model, which implies conditions for the gauge and Yukawa couplings.
The so defined boundary value problem must be solved numerically until
a set of model parameters has been found consistent with all
user-supplied boundary conditions.

\fs is a \mathematica and C++ package designed to create a fast and easily
adaptable spectrum generator in C++ for any SUSY model.
The user specifies the model by giving the
superfield content, superpotential, gauge symmetries and mass mixings
in form of \sarah model files.  The boundary conditions on the model
parameters must be specified in a separate \code{FlexibleSUSY.m.in} file.
Based on this information \fs uses \sarah to obtain tree-level
expressions for the mass matrices and electroweak symmetry breaking
conditions, one-loop self energies, one-loop tadpoles corrections and
two-loop renormalization group equations (RGEs) for the model.
Additional corrections which have been calculated elsewhere, such as
two-loop corrections to the Higgs masses\footnote{By default
  \fs has two-loop corrections to the Higgs masses for the
  MSSM
  \cite{Degrassi:2001yf,Brignole:2001jy,Dedes:2002dy,Brignole:2002bz,Dedes:2003km}
  and NMSSM \cite{Degrassi:2009yq} in FORTRAN files supplied by Pietro
  Slavich. These are the same corrections which are implemented in
  many of the public spectrum generators.} may be added by the user.
These algebraic expressions are converted into C++ code and are put
into classes with well-defined interfaces to allow for easy exchange,
extension and reuse of the modules.  All of these classes are finally
combined to a complete spectrum generator, which solves the
user-defined boundary value problem.  For this task \fs uses some
parts of Softsusy \cite{Allanach:2001kg}, the very fast Eigen library
\cite{eigen}, augmented by LAPACK, as well as the GNU scientific
library and the Boost library to create numerical routines which solve
the RGEs and boundary conditions simultaneously.  If a solution has
been found the pole mass spectrum is eventually calculated using full
one-loop self-energies (and leading two-loop Higgs self-energy
contributions for the MSSM and NMSSM).

The standard input and output of the generated spectrum generator is
the SLHA format \cite{Skands:2003cj,Allanach:2008qq}, which is
intended for the communication between MSSM and NMSSM spectrum
generators and observable calculators.  The user has control over the
SLHA block names by editing the SARAH model files and may also add
extra blocks in the \code{FlexibleSUSY.m} model file, as described in
\secref{sec:modfile}.
Internally, \fs uses the SLHAea library \cite{slhaea} to read and
write the SLHA files.  The internally stored SLHAea object can be
passed to other programs at the C++ level for inter program
communication.
We have tested that the HDECAY $3.4$ \cite{hdecay} and SDECAY $1.4$
\cite{sdecay}, which are shipped with SUSY-HIT \cite{susyhit},
understand the SLHA output of \fs and give matching output to that
from Softsusy.
For models which go beyond the MSSM and NMSSM no fixed standard exists
which specifies the spectrum generator input and output format.  For
this reason, \fs allows the user to control the SLHA input and output
blocks in order to simplify the process of passing the output of a
custom \fs-generated spectrum generator to any private or new public
tool developed for that model.

\subsection{Design goals}

Since the calculation of the pole mass spectrum in a SUSY model is a
non-trivial task, \fs is designed with the following points in mind:

\paragraph{Modularity}

The large variety of supersymmetric models and potential
investigations makes it likely that the user wants to modify the
generated spectrum generator source code or reuse components in
further programs.  \fs offers two levels to influence the code: (i) On
the \mathematica model file level the model itself or GUT/weak scale
boundary conditions as well as input and output parameters can be
controlled (see \secref{sec:quick-start-alternative-models} and
\secref{sec:adapting-model-files} for examples).  (ii) In particular
\fs uses C++ object orientation features to modularize the source code
so that it is sharply divided into building blocks performing
distinct duties.
This modular architecture makes it easy
for the user to modify, reuse, replace or
extend the individual components (see \secref{sec:adapting-cpp-code}
for examples).  An important application of this concept are the
boundary conditions, for which the C++ level offers a wider range of
possibilities.  The boundary conditions solver provides a plugin
mechanism via a common \code{Constraint} interface, which allows a
user to exchange or add boundary conditions at any scale.
To realize this, all (derived) constraint objects are intentionally
kept outside the solver.  Despite being independent of one another,
they can fit together with the aid of class inheritance.  An
elaborate example of a tower of effective field theories and multiple
matching scales is presented in \secref{sec:tower construction}.
Alternatively, the modular structure makes it straightforward to take
\fs generated code for e.g.\ RGEs or self-energies and reuse it in an
existing code for some other purpose.  Conversely, it is also easy to
include code from elsewhere into the spectrum generator.  For an
example see \secref{sec:integrating-custom-built}.

\paragraph{Speed}

Exploring the parameter space of supersymmetric models with a high
number of free parameters is quite time consuming.  Therefore \fs aims
to produce spectrum generators with a short run-time.  The two most
time consuming parts of a SUSY spectrum generator are usually the
calculation of the two-loop $\beta$-functions and the pole masses of
mixed particles:
\begin{itemize}
\item \emph{Calculation of the $\beta$-functions:} The RG solving
  algorithms usually need $O(10)$ iterations between the high and the
  low scale to find a set of parameters consistent with all boundary
  conditions with a $0.01\%$ precision goal.  During each iteration
  the Runge-Kutta algorithm needs to calculate all $\beta$-functions
  $O(50)$ times.  Most two-loop $\beta$-functions involve $O(50)$
  matrix multiplications and additions.  All together one arrives at
  $O(10^4)$ matrix operations.  To optimize these, \fs uses the fast
  linear algebra package \href{Eigen}{http://eigen.tuxfamily.org}.
  Eigen uses C++ expression templates to remove temporary objects and
  enable lazy evaluation of the expressions.  It supports explicit
  vectorization, and provides fixed-size matrices to avoid dynamic
  memory allocation.  All of these features in combination result in
  very fast code for the calculation of the $\beta$-functions in \fs.
\item \emph{Calculation of the pole masses:} The second most time
  consuming part is the precise calculation of the pole masses for
  mixed particles.  For each particle $\psi_k$ in a multiplet the full
  self-energy matrix $\Sigma^\psi_{ij}(p=m^\text{tree}_{\psi_k})$ has
  to be evaluated.  Each self-energy matrix entry again involves the
  calculation of $O(50)$ Feynman diagrams, each involving the
  calculation of vertices and a loop-function.  All in all, one
  arrives at $O(500)$ Feynman diagrams and $O(10^4)$ loop function
  evaluations.  To speed up the calculation of the pole masses \fs
  makes use of multi-threading, where each pole mass is calculated in
  a separate thread.  This allows the operating system to distribute
  these calculations among different CPU cores.  With this technique
  one can gain a speed-up of $20$--$30\%$.
\end{itemize}

\paragraph{Alternative boundary value problem solvers}

Furthermore, the standard algorithm which solves the user-defined
boundary value problem via a fixed-point iteration is not guaranteed
to converge in all regions of the model parameter space.  Therefore,
\fs has been intentionally designed to allow for alternative solvers
to search for solutions in such critical parameter regions.  A
subsequent release with an alternative solver is already planned.

\paragraph{Towers of effective theories}

In \fs the standard fixed-point iteration solver has been generalized
to handle towers of models (effective theories), which are matched at
intermediate scales.  An example of such a tower construction will be
given in \secref{sec:tower construction}, where right-handed neutrinos
are integrated out at the see-saw scale, between the SUSY and the GUT
scale.

\subsection{Current limitations and future extensions}
Although we try to handle as many models as possible, there are still
some limitations to what can be done. In its current form \fs assumes
all couplings are real, therefore it is limited to $CP$-conserving
versions of SUSY models.  Although it is implicit in the title, we
would like to stress that currently we cannot provide a spectrum
generator for non-SUSY models.  This also means that like other
spectrum generators it is difficult to reliably predict the mass
spectrum in extreme cases such as Split-SUSY
\cite{Wells:2003tf,ArkaniHamed:2004fb,Giudice:2004tc}, where the mass
scales of the SUSY model are drastically split, leading to very large
logarithms which are not resummed.  In such cases the tower of
effective theories offers the best possibility for a solution.
However, for this to work \fs must be extended to include non-SUSY
models.  Finally, the gauge group of the model is currently restricted
to be semi-simple and to contain the Standard Model gauge group as
factor, so that the SUSY model can be directly matched to the SM at
low energies.
Future releases which extend \fs beyond each of these limitations are
already planned.  

\section{Quick start}
\label{sec:download}

\subsection{Requirements}

\fs can be downloaded from \url{http://flexiblesusy.hepforge.org}.  To
create a custom spectrum generator the following requirements are
necessary:
\begin{itemize}
\item \mathematica, version 7 or higher
\item SARAH, version 4.0.4 or higher \url{http://sarah.hepforge.org}
\item C++11 compatible compiler (g++ 4.4.7 or higher, clang++ 3.1 or
  higher, icpc 12.1 or higher)
\item FORTRAN compiler (gfortran, ifort etc.)
\item Eigen library, version 3.1 or higher
  \url{http://eigen.tuxfamily.org}
\item Boost library, version 1.37.0 or higher
  \url{http://www.boost.org}
\item GNU scientific library \url{http://www.gnu.org/software/gsl}
\item an implementation of LAPACK \url{http://www.netlib.org/lapack}
  such as ATLAS \url{http://math-atlas.sourceforge.net} or
  Intel Math Kernel Library \url{http://software.intel.com/intel-mkl}
\end{itemize}
Optional:
\begin{itemize}
\item Looptools, version 2.8 or higher
  \url{http://www.feynarts.de/looptools}
\end{itemize}

\subsection{Downloading \fs and generating a first spectrum generator}
\label{sec:quick-start-cmssm}

\fs can be downloaded as a gzipped tar file from
\url{http://flexiblesusy.hepforge.org}.  To download and install
version 1.0.3 run:
\begin{lstlisting}[language=bash]
$ wget https://www.hepforge.org/archive/flexiblesusy/FlexibleSUSY-1.0.3.tar.gz
$ tar -xf FlexibleSUSY-1.0.3.tar.gz
$ cd FlexibleSUSY-1.0.3
\end{lstlisting}
A CMSSM spectrum generator can be created with the following three
commands:
\begin{lstlisting}[language=bash]
$ ./createmodel --name=MSSM
$ ./configure --with-models=MSSM
$ make
\end{lstlisting}
The first command creates the model directory \code{models/MSSM/}
together with a CMSSM model file accompanied by
a specimen SLHA input file.  The \code{configure} script checks
the system requirements and creates the \code{Makefile}.  See
\code{./configure --help} for more options.  Executing \code{make}
will start \mathematica to generate the spectrum generator and compile
it.  The resulting executable can be run like this:
\begin{lstlisting}[language=bash]
$ cd models/MSSM 
$ ./run_MSSM.x --slha-input-file=LesHouches.in.MSSM
\end{lstlisting}
When executed, the spectrum generator tries to find a set of \DRbar\
model parameters consistent with all CMSSM boundary conditions for the
parameter point given in the SLHA input file
\code{LesHouches.in.MSSM}.  Afterwards, the pole mass
spectrum and mixing matrices are calculated and written to the standard
output in SLHA format \cite{Skands:2003cj,Allanach:2008qq}.  For the
parameter point given in the above example the calculated pole mass
spectrum reads
\begin{lstlisting}
Block MASS
   1000021     1.15236966E+03   # Glu
   1000024     3.85774334E+02   # Cha_1
   1000037     6.50460073E+02   # Cha_2
        25     1.14766149E+02   # hh_1
        35     7.06792640E+02   # hh_2
        37     7.11388516E+02   # Hpm_2
        36     7.06523105E+02   # Ah_2
   1000012     3.51856376E+02   # Sv_1
   1000014     3.53042556E+02   # Sv_2
   1000016     3.53046504E+02   # Sv_3
   1000022     2.03889780E+02   # Chi_1
   1000023     3.85760714E+02   # Chi_2
   1000025     6.36544884E+02   # Chi_3
   1000035     6.50133768E+02   # Chi_4
   1000001     9.66656018E+02   # Sd_1
   1000003     1.00983181E+03   # Sd_2
   1000005     1.01651873E+03   # Sd_3
   2000001     1.01653005E+03   # Sd_4
   2000003     1.06089534E+03   # Sd_5
   2000005     1.06090238E+03   # Sd_6
   1000011     2.22570305E+02   # Se_1
   1000013     2.29864536E+02   # Se_2
   1000015     2.29888846E+02   # Se_3
   2000011     3.61946671E+02   # Se_4
   2000013     3.61950866E+02   # Se_5
   2000015     3.63136031E+02   # Se_6
   1000002     8.09787818E+02   # Su_1
   1000004     1.01454197E+03   # Su_2
   1000006     1.01981109E+03   # Su_3
   2000002     1.02015269E+03   # Su_4
   2000004     1.05807759E+03   # Su_5
   2000006     1.05808168E+03   # Su_6
\end{lstlisting}

\subsection{Spectrum generators for alternative models}
\label{sec:quick-start-alternative-models}

\fs already comes with plenty of predefined models: the CMSSM (simply
called \code{MSSM}), the NMSSM in its $Z_3$-symmetric form (called \code{NMSSM}), $Z_3$-violating NMSSM (\code{SMSSM}), the USSM (\code{UMSSM}),
the NUHM \ESSM (\code{E6SSM}) \cite{Athron:2007en}, the
right-handed neutrino extended MSSM (\code{MSSMRHN}), the NUHM-MSSM
(\code{NUHMSSM}) and the $R$-symmetric MSSM (\code{MRSSM})
\cite{Kribs:2007ac}.  See the content of \code{model_files/} for all
predefined model files.  For all these models spectrum generators can
be generated easily like for the CMSSM in
\secref{sec:quick-start-cmssm}.  The spectrum generator for the
$Z_3$-symmetric NMSSM for example can be generated like this:
\begin{lstlisting}[language=bash]
$ ./createmodel --name=NMSSM
$ ./configure --with-models=NMSSM
$ make
\end{lstlisting}
One of the design goals is modularity and the possibility to easily
construct custom spectrum generators.  The details of the
customization can be found in Sections
\ref{sec:modfile}--\ref{sec:Flexible}.  As a simple example consider
the NMSSM.  The NMSSM variant above unifies all soft-breaking
trilinear scalar couplings at the GUT scale.  In order to relax this
constraint and use a separate value for $A_\lambda$ at the GUT scale
one can edit the model file \code{model_files/NMSSM/FlexibleSUSY.m.in} and
change the lines
\begin{lstlisting}[language=Mathematica]
EXTPAR = { {61, LambdaInput} };

HighScaleInput = {
   ...
   {T[\[Lambda]], Azero LambdaInput},
   ...
};
\end{lstlisting}
into
\begin{lstlisting}[language=Mathematica]
EXTPAR = { {61, LambdaInput},
           {63, ALambdaInput} };

HighScaleInput = {
   ...
   {T[\[Lambda]], ALambdaInput LambdaInput},
   ...
};
\end{lstlisting}
The value of $A_\lambda$ at the GUT scale can then be set in the SLHA
input file in the \code{EXTPAR} block entry $63$ via
\begin{lstlisting}
Block EXTPAR
   61   0.1                  # LambdaInput
   63   -100                 # ALambdaInput
\end{lstlisting}

\section{Setting up a FlexibleSUSY model}
\label{sec:modfile}

A general (non-constrained) softly broken SUSY model is defined by the
gauge group, the field content and mixings as well as the
superpotential and the soft-breaking Lagrangian.  In order to create a
spectrum generator for such a SUSY model with \fs, the aforementioned
model properties have to be defined in a SARAH model file.  The SARAH
model file can be put into the \code{sarah/<model>/} directory.  See
the SARAH manual \cite{Staub:2008uz,Staub:2013tta} for a detailed
explanation of how to write such a model file.  Note that SARAH
already is distributed with a lot of predefined models, which can be used with
\fs immediately.

The model boundary conditions are defined in the \fs model file
\code{FlexibleSUSY.m}, which has to be located in the model directory
\code{models/<model>/}.  To add this the user should create a
\code{FlexibleSUSY.m.in} file in the directory
\code{model_files/<model>/}.  When the \code{./createmodel} script is
executed, the \code{FlexibleSUSY.m} file is created from the
\code{model_files/<model-file-name>/FlexibleSUSY.m.in} file, where the
directory \code{<model-file-name>} is specified by the
\code{--model-file=<model-file-name>} option.  If no such option is
given the directory matching the \code{--name=<model>} option is used.
In either case the \code{FlexibleSUSY.m} file which is created is then
automatically placed in the directory \code{models/<model>/}.  Note
that many predefined example model files can already be found in
\code{model_files/}.

In the following it is explained how the
boundary conditions can be defined on the basis of the CMSSM.  The
application to other models is straightforward.  The CMSSM model file
reads:
\begin{lstlisting}[language=Mathematica]
FSModelName = "@CLASSNAME@";

MINPAR = {
   {1, m0},
   {2, m12},
   {3, TanBeta},
   {4, Sign[\[Mu]]},
   {5, Azero}
};

EWSBOutputParameters = { B[\[Mu]], \[Mu] };

HighScale = g1 == g2;

HighScaleFirstGuess = 2.0 10^16;

HighScaleMinimum = 1.0 10^10; (* optional *)

HighScaleMaximum = 1.0 10^18; (* optional *)

HighScaleInput = {
   {T[Ye], Azero*Ye},
   {T[Yd], Azero*Yd},
   {T[Yu], Azero*Yu},
   {mHd2, m0^2},
   {mHu2, m0^2},
   {mq2, UNITMATRIX[3] m0^2},
   {ml2, UNITMATRIX[3] m0^2},
   {md2, UNITMATRIX[3] m0^2},
   {mu2, UNITMATRIX[3] m0^2},
   {me2, UNITMATRIX[3] m0^2},
   {MassB, m12},
   {MassWB, m12},
   {MassG, m12}
};

SUSYScale = Sqrt[Product[M[Su[i]]^(Abs[ZU[i,3]]^2 + Abs[ZU[i,6]]^2), {i,6}]];

SUSYScaleFirstGuess = Sqrt[m0^2 + 4 m12^2];

SUSYScaleInput = {};

LowScale = SM[MZ];

LowScaleFirstGuess = SM[MZ];

LowScaleInput = {
   {Yu, Automatic},
   {Yd, Automatic},
   {Ye, Automatic},
   {vd, 2 MZDRbar / Sqrt[GUTNormalization[g1]^2 g1^2 + g2^2]
           Cos[ArcTan[TanBeta]]},
   {vu, 2 MZDRbar / Sqrt[GUTNormalization[g1]^2 g1^2 + g2^2]
           Sin[ArcTan[TanBeta]]}
};

InitialGuessAtLowScale = {
   {vd, SM[vev] Cos[ArcTan[TanBeta]]},
   {vu, SM[vev] Sin[ArcTan[TanBeta]]},
   {Yu, Automatic},
   {Yd, Automatic},
   {Ye, Automatic}
};

InitialGuessAtHighScale = {
   {\[Mu]   , 1.0},
   {B[\[Mu]], 0.0}
};

UseHiggs2LoopMSSM = True;
EffectiveMu = \[Mu];

OnlyLowEnergyFlexibleSUSY = False; (* default *)

PotentialLSPParticles = { Chi, Cha, Glu, Sv, Su, Sd, Se };

DefaultPoleMassPrecision = MediumPrecision;
HighPoleMassPrecision    = {hh, Ah, Hpm};
MediumPoleMassPrecision  = {};
LowPoleMassPrecision     = {};

(* optional *)
ExtraSLHAOutputBlocks = {
   {ALPHA, {{ArcCos[Pole[ZH[1,2]]]}}},
   {HMIX , {{1, \[Mu]},
            {2, vu / vd},
            {3, Sqrt[vu^2 + vd^2]},
            {4, M[Ah[2]]^2},
            {101, B[\[Mu]]},
            {102, vd},
            {103, vu} } },
   {Au,    {{1, 1, T[Yu][1,1] / Yu[1,1]},
            {2, 2, T[Yu][2,2] / Yu[2,2]},
            {3, 3, T[Yu][3,3] / Yu[3,3]} } },
   {Ad,    {{1, 1, T[Yd][1,1] / Yd[1,1]},
            {2, 2, T[Yd][2,2] / Yd[2,2]},
            {3, 3, T[Yd][3,3] / Yd[3,3]} } },
   {Ae,    {{1, 1, T[Ye][1,1] / Ye[1,1]},
            {2, 2, T[Ye][2,2] / Ye[2,2]},
            {3, 3, T[Ye][3,3] / Ye[3,3]} } },
   {MSOFT, {{1, MassB},
            {2, MassWB},
            {3, MassG},
            {21, mHd2},
            {22, mHu2},
            {31, Sqrt[ml2[1,1]]},
            {32, Sqrt[ml2[2,2]]},
            {33, Sqrt[ml2[3,3]]},
            {34, Sqrt[me2[1,1]]},
            {35, Sqrt[me2[2,2]]},
            {36, Sqrt[me2[3,3]]},
            {41, Sqrt[mq2[1,1]]},
            {42, Sqrt[mq2[2,2]]},
            {43, Sqrt[mq2[3,3]]},
            {44, Sqrt[mu2[1,1]]},
            {45, Sqrt[mu2[2,2]]},
            {46, Sqrt[mu2[3,3]]},
            {47, Sqrt[md2[1,1]]},
            {48, Sqrt[md2[2,2]]},
            {49, Sqrt[md2[3,3]]} } }
};
\end{lstlisting}
The first line \code{FSModelName = "@CLASSNAME@";} will be replaced
with \code{FSModelName = "<model>";} in the generated
\code{FlexibleSUSY.m} file, where \code{<model>} is specified by the
\code{--name=<model>} option for the \code{./createmodel} script.  So
the variable \code{FSModelName} then contains the name of the \fs
model.

All non-Standard Model input variables must be specified in the lists
\code{MINPAR} and \code{EXTPAR}.  These two variables refer to the
MINPAR and EXTPAR blocks in a SLHA input file \cite{Skands:2003cj}.
The list elements are two-component lists where the first entry is the
SLHA index in the MINPAR or EXTPAR block, respectively, and the second
entry is the name of the input parameter.  In the above example the
input parameters are the universal soft-breaking parameters $m_0$,
$M_{1/2}$, $A_0$ as well as $\tan\beta$ and $\sign\mu$.

Using the variable \code{EWSBOutputParameters} the user can
specify the model parameters that are output of the electroweak
symmetry breaking consistency conditions.  When imposing the EWSB, \fs
will adjust these parameters until the EWSB conditions are fulfilled.
In the CMSSM example above these are the superpotential parameter
$\mu$ and its corresponding soft-breaking parameter $B\mu$.  In the
NMSSM the parameters $\kappa$, $|v_s|$ and $m_s^2$ are usually chosen
for this purpose.

Furthermore, the user has to specify three model constraints:
low-scale, SUSY-scale and high-scale.  In \fs they are named as
\code{LowScale}, \code{SUSYScale} and \code{HighScale}.  For each
constraint there is (i) a scale definition (named after the
constraint), (ii) an initial guess for the scale (concatenation of the
constraint name and \code{FirstGuess}) and (iii) a list of parameter
settings to be applied at the scale (concatenation of the constraint
name and \code{Input}).  Optionally a minimum and a maximum value for
the scale can be given (concatenation of the constraint name and
\code{Minimum} or \code{Maximum}, respectively).  The latter avoids
underflows or overflows of the scale value during the iteration.  This
is especially useful in models where the iteration is very unstable
and the value of the scale is very sensitive to the model parameters.
The meaning of the three constraints is the following:
\begin{itemize}
\item \emph{High-scale constraint:} The high-scale constraint is
  usually the GUT-scale constraint, imposed at the scale where the
  gauge couplings $g_1$ and $g_2$ unify.  The high-scale can be
  defined by an equation of the form \code{g1 == g2} or by a fixed
  numerical value.  Note that \fs GUT-normalizes all gauge couplings.
  Thus, the high-scale definition takes the simple form \code{g1 ==
    g2}.  As a consequence in the calculation of the VEVs $v_u$ and
  $v_d$ from $M_Z$ and $\tan\beta$ at the low-scale the
  GUT-normalization has to be taken into account, see the example
  above.
\item \emph{SUSY-scale constraint:} The SUSY-scale is the typical mass
  scale of the SUSY particle spectrum.  At this scale \fs imposes the
  EWSB conditions and calculates the pole mass spectrum.  The
  SUSY-scale, $M_S$, is defined as
  \begin{align}
    M_S = \sqrt{\prod_{i=1}^6 m_{\tilde{u}_i}^{|(Z_u)_{i3}|^2 + |(Z_u)_{i6}|^2}} ,
    \label{eq:definition_of_MS}
  \end{align}
  where $m_{\tilde{u}_i}$ is the \DRbar\ mass of the $i$th up-type
  squark and $Z_u$ is the up-type squark mixing matrix.  The
  definition \eqref{eq:definition_of_MS} is equivalent to the usual
  choice $M_S = \sqrt{m_{\tilde{t}_1}m_{\tilde{t}_2}}$ without squark
  flavour mixing.
\item \emph{Low-scale constraint:} The low-scale constraint is the
  constraint where the SUSY model is matched to the Standard Model.
  This is done by automatically calculating the gauge couplings $g_i$
  ($i=1,2,3$) of the SUSY model from the known Standard Model
  quantities $\alpha_{\text{e.m.}}(M_Z)$, $\alpha_{s}(M_Z)$, $M_Z$,
  $M_W$.  The details of this calculation are explained in
  \secref{sec:calculation-of-gauge-couplings}.  Currently this scale
  is fixed to be the $Z$ pole mass scale $M_Z$.  Optionally the Yukawa
  couplings $y_f$ ($f=u,d,e$) can be calculated automatically from the
  known Standard Model fermion masses $m_f$ by setting their values to
  \code{Automatic}.  This automatic calculation is explained in
  \secref{sec:calculation-of-yukawa-couplings}.
\end{itemize}
The variables \code{LowScaleInput}, \code{SUSYScaleInput} and
\code{HighScaleInput}, which list the parameter settings for imposing
the constraints can contain as elements any of the following:
\begin{itemize}
\item Two-component lists of the form \code{\{parameter, value\}},
  which indicates that the \code{parameter} is set to \code{value} at
  the defined scale.  If the \code{value} should be read from the SLHA
  input file, it must be written as \code{LHInput[value]}.  Example:
  \begin{lstlisting}
SUSYScaleInput = {
   {mHd2, m0^2},
   {mHu2, LHInput[mHu2]}
};
  \end{lstlisting}
  In this example the parameter \code{mHd2} is set to the value of
  \code{m0^2}, and \code{mHu2} is set to the value given in the SLHA
  input file in block \code{MSOFTIN}, entry 22 at the SUSY scale.  The
  SLHA block names and keys for the MSSM and NMSSM are defined in
  SARAH's \code{parameters.m} file, see the SARAH manual or
  \cite{Staub:2010jh}.  For the Standard Model Yukawa couplings
  \code{Yu}, \code{Yd}, \code{Ye} the value \code{Automatic} is
  allowed, which triggers their automatic determination from the known
  Standard Model quark and lepton masses, see
  \secref{sec:calculation-of-yukawa-couplings}.

\item The function \code{FSMinimize[parameters, function]} can be
  given, where \code{parameters} is a list of model parameters and
  \code{function} is a function of these parameters.
  \code{FSMinimize[parameters, function]} will numerically vary the
  \code{parameters} until the \code{function} is minimized.  Example:
  \begin{lstlisting}
FSMinimize[{vd,vu},
           (SM[MZ] - Pole[M[VZ]])^2 / STANDARDDEVIATION[MZ]^2 +
           (SM[MH] - Pole[M[hh[1]]])^2 / STANDARDDEVIATION[MH]^2]
  \end{lstlisting}
  Here, the parameters \code{vu} and \code{vd} are varied until the
  function
  \begin{align}
    \chi^2(v_d,v_u) =
    \frac{(\texttt{SM[MZ]}-m_Z^\pole)^2}{\sigma_{m_Z}^2} +
    \frac{(\texttt{SM[MH]}-m_{h_1}^\pole)^2}{\sigma_{m_h}^2}
  \end{align}
  is minimal.  The constants \code{SM[MZ]}, \code{SM[MH]},
  $\sigma_{m_Z}$ and $\sigma_{m_h}$ are defined in
  \code{src/ew_input.hpp} to be
  \begin{align}
    \texttt{SM[MZ]} &= 91.1876, &
    \texttt{SM[MH]} &= 125.9, \\
    \sigma_{m_Z} &= 0.0021, &
    \sigma_{m_h} &= 0.4 .
  \end{align}

\item The function \code{FSFindRoot[parameters, functions]} can be
  given, where \code{parameters} is a list of model parameters and
  \code{functions} is a list of functions of these parameters.
  \code{FSFindRoot[parameters, functions]} will numerical vary the
  \code{parameters} until the \code{functions} are zero.  Example:
  \begin{lstlisting}
FSFindRoot[{vd,vu},
           {SM[MZ] - Pole[M[VZ]], SM[MH] - Pole[M[hh[1]]]}]
  \end{lstlisting}
  Here, the parameters \code{vu} and \code{vd} are varied until the
  vector-valued function
  \begin{align}
    f(v_d,v_u) =
    \begin{pmatrix}
      \texttt{SM[MZ]} - m_Z^\pole \\
      \texttt{SM[MH]} - m_{h_1}^\pole
    \end{pmatrix}
  \end{align}
  is zero.
\end{itemize}
Finally, the user can set an initial guess for the model parameters at
the low- and high-scale using the variables
\code{InitialGuessAtLowScale} and \code{InitialGuessAtHighScale},
respectively.  The gauge couplings will be guessed automatically at
the low-scale from the known Standard Model parameters.

\fs allows the user to add leading two-loop contributions to the
$CP$-even and $CP$-odd Higgs self-energies as well as to the $CP$-even Higgs
tadpoles.  For MSSM-like models (with two $CP$-even Higgs bosons, one
$CP$-odd Higgs boson, one neutral Goldstone boson) routines for
calculating these corrections will be generated by setting
\code{UseHiggs2LoopMSSM = True} in the model file and by defining the
effective $\mu$-term \code{EffectiveMu = \\[Mu]}.  This will add the
zero-momentum corrections of the order $O(y_t^4 + y_b^2 y_t^2 +
y_b^4)$, $O(y_t^2 g_3^2)$, $O(y_b^2 g_3^2)$, $O(y_\tau^4)$ from
\cite{Degrassi:2001yf,Brignole:2001jy,Dedes:2002dy,Brignole:2002bz,Dedes:2003km}.
For NMSSM-like models (with three $CP$-even Higgs bosons, two $CP$-odd
Higgs bosons, one neutral Goldstone boson) the two-loop contributions
are generated by setting \code{UseHiggs2LoopNMSSM = True} and by
defining the effective $\mu$-term like \code{EffectiveMu = \\[Lambda]
  vS / Sqrt[2]}, for example.  This will add the zero-momentum
corrections of the order $O(y_t^2 g_3^2)$, $O(y_b^2 g_3^2)$ from
\cite{Degrassi:2009yq} and also the MSSM contributions of the order
$O(y_\tau^4)$, $O(y_t^4 + y_t^2 y_b^2 + y_b^4)$ as well
\cite{Brignole:2001jy,Dedes:2003km} which only represent a partial
correction for that order in the NMSSM, but can be a good
approximation when singlet mixing is very small\footnote{These
  corrections may be disabled in the SLHA file, as described in
  section \secref{sec:PoleMasses}.}.

 The corrections can then be used in the calculation of the Higgs
 masses, when appropriate settings are selected in the SLHA file. Note
 that even in the NMSSM the user must make an important decision as
 to whether or not to enable the generated MSSM corrections which are
 incomplete in the NMSSM.  We feel that it is valuable to have
 these MSSM corrections for scenarios where singlet mixing is
 very small and in particular for cross checks against the MSSM when
 close to the MSSM limit of the model.  However in cases where the
 singlet mixing is large the result at $O(y_t^4 + y_t^2 y_b^2 +
 y_b^4)$ and $O(y_\tau^4)$ will not be complete and including these
 corrections could in principle even make the numerical result further
 away from the correct two loop result at that order if there is a
 cancellation with the missing contributions.  So while including such
 partial two-loop corrections does not change the formal accuracy
 the user should choose whether or not to include these corrections
 based on the physics they study.  Similarly in models that go beyond
 the NMSSM the user must decide based on the physics whether or not these
 corrections will give the leading two-loop corrections in their
 model.

One can create a pure low-energy model by setting
\code{OnlyLowEnergyFlexibleSUSY = True}.  In this case the high-scale
constraint is ignored and only the low-scale and SUSY-scale
constraints are used.  All model parameters which are not specified in
\code{MINPAR} or \code{EXTPAR} will then be read from the
corresponding input blocks in the SLHA input file and will be set at
the SUSY-scale.  An example of such a pure low-energy model is the
MRSSM, where the three gauge couplings do not unify at a common scale.

\fs can create the helper function \code{get_lsp()}, which finds the
lightest supersymmetric particle (LSP).  To have this function be
created the model file variable \code{PotentialLSPParticles} must be
set to a list of SUSY particles which are potential LSPs.  In the
model file example above, the particles \code{Chi}, \code{Cha},
\code{Glu}, \code{Sv}, \code{Su}, \code{Sd}, \code{Se} (neutralino,
chargino, gluino, sneutrino, up-type squark, down-type squark,
selectron) are considered to be LSP candidates.

Finally, with the variable \code{ExtraSLHAOutputBlocks} the user can
define extra SLHA output blocks.  The values of the block entries will
be calculated at the output scale, which is specified in entry $12$ in
the SLHA input block \code{MODSEL}.  In the example model file above
the following six extra SLHA compliant output blocks are defined:
\code{ALPHA}, \code{HMIX}, \code{Au}, \code{Ad}, \code{Ae} and
\code{MSOFT}.  The \code{ALPHA} output block contains the $CP$-even
Higgs pole mass mixing angle as the only entry.  The \code{HMIX}
output block contains the $\mu$-parameter, the ratio $\tan\beta =
v_u/v_d$, the combination $v=\sqrt{v_u^2 + v_d^2}$, the squared mass
of the $CP$-odd Higgs $m_A^2$, the soft-breaking parameter $B\mu$ and
the values of $v_u$ and $v_d$, all defined in the \DRbar\ scheme.  In
an analogous way four more output blocks for the soft-breaking \DRbar\
parameters are defined.  For a CMSSM example parameter point with $m_0
= 125\unit{GeV}$, $M_{1/2} = 500\unit{GeV}$, $\tan\beta = 10$,
$\sign\mu = +1$ and $A_0 = 0$ the \fs-generated CMSSM spectrum
generator writes the so defined blocks to the output in the form
\begin{lstlisting}
Block ALPHA Q= 8.76740936E+02
           1.06784138E-01   # ArcCos(Pole(ZH(0,1)))
Block HMIX Q= 8.76740936E+02
     1     6.31218384E+02   # Mu
     2     9.67312621E+00   # vu/vd
     3     2.44053433E+02   # Sqrt(Sqr(vd) + Sqr(vu))
     4     5.36777230E+05   # Sqr(MAh(1))
   101     5.49048159E+04   # BMu
   102     2.50962986E+01   # vd
   103     2.42759663E+02   # vu
Block Au Q= 8.76740936E+02
  1  1    -1.14477419E+03   # TYu(0,0)/Yu(0,0)
  2  2    -1.14476911E+03   # TYu(1,1)/Yu(1,1)
  3  3    -8.83902977E+02   # TYu(2,2)/Yu(2,2)
Block Ad Q= 8.76740936E+02
  1  1    -1.40026447E+03   # TYd(0,0)/Yd(0,0)
  2  2    -1.40025976E+03   # TYd(1,1)/Yd(1,1)
  3  3    -1.30885006E+03   # TYd(2,2)/Yd(2,2)
Block Ae Q= 8.76740936E+02
  1  1    -3.00005426E+02   # TYe(0,0)/Ye(0,0)
  2  2    -3.00000006E+02   # TYe(1,1)/Ye(1,1)
  3  3    -2.98364373E+02   # TYe(2,2)/Ye(2,2)
Block MSOFT Q= 8.76740936E+02
     1     2.09018579E+02   # MassB
     2     3.88257873E+02   # MassWB
     3     1.11544211E+03   # MassG
    21     1.09683411E+05   # mHd2
    22    -3.85898988E+05   # mHu2
    31     3.54416224E+02   # Sqrt(ml2(0,0))
    32     3.54412891E+02   # Sqrt(ml2(1,1))
    33     3.53407333E+02   # Sqrt(ml2(2,2))
    34     2.22035720E+02   # Sqrt(me2(0,0))
    35     2.22024873E+02   # Sqrt(me2(1,1))
    36     2.18731935E+02   # Sqrt(me2(2,2))
    41     1.02053367E+03   # Sqrt(mq2(0,0))
    42     1.02053107E+03   # Sqrt(mq2(1,1))
    43     9.40760849E+02   # Sqrt(mq2(2,2))
    44     9.82565930E+02   # Sqrt(mu2(0,0))
    45     9.82563167E+02   # Sqrt(mu2(1,1))
    46     8.09126982E+02   # Sqrt(mu2(2,2))
    47     9.77979392E+02   # Sqrt(md2(0,0))
    48     9.77976666E+02   # Sqrt(md2(1,1))
    49     9.73121951E+02   # Sqrt(md2(2,2))
\end{lstlisting}

\section{Structure of the spectrum generator}
\label{sec:SpecGenStruct}

In this section we explain the internals of \fs's automatically
generated spectrum generator.

As mentioned in \secref{sec:Program}, \fs uses SARAH-generated
expressions for the $\beta$-functions, mass matrices, self-energies
and EWSB conditions plus the user-defined parameter boundary
conditions to create a spectrum generator in C++.  This program takes
the Standard Model and user-defined input parameters and numerically
solves the boundary value problem, which is defined by the RG
equations and the boundary conditions.  If a solution is found the
pole mass spectrum is calculated.

In the following it is explained how this procedure is realized in
\fs.  As mentioned in \secref{sec:Program} one of \fs's design goals
is to create modular C++ code to allow for an easy exchange, extension
and reuse of the generated modules.  For this reason
\secref{sec:ModelParametersAndRGEs} first of all briefly describes the
so-called C++ ``model class'' hierarchy, which contains the general
model information, such as parameters, $\beta$-functions, \DRbar\ mass
spectrum, EWSB, self-energies, and the pole mass spectrum.
\secref{sec:boundary-conditions} describes how boundary conditions on
the model parameters are implemented in general at the C++ level.
Subsections \ref{sec:calculation-of-gauge-couplings}--\ref{sec:ewsb}
then show the two concrete boundary conditions, which are always
imposed: The matching of the model parameters to the Standard Model
and the electroweak symmetry breaking.
In \secref{sec:TreeLevelSpectrum} we describe the conventions used to
calculate the \DRbar\ mass spectrum given a set of \DRbar\ model
parameters.
Afterwards, in \secref{sec:TwoScaleFixedPointIteration} the algorithm,
which solves the user-defined boundary value problem is described on
the basis of the CMSSM example given in \secref{sec:modfile}.
Finally, \secref{sec:PoleMasses} explains how the pole mass spectrum
is obtained from the \DRbar\ model parameters after a solution to the
boundary value problem has been found.

\subsection{Model parameters and RGEs}
\label{sec:ModelParametersAndRGEs}

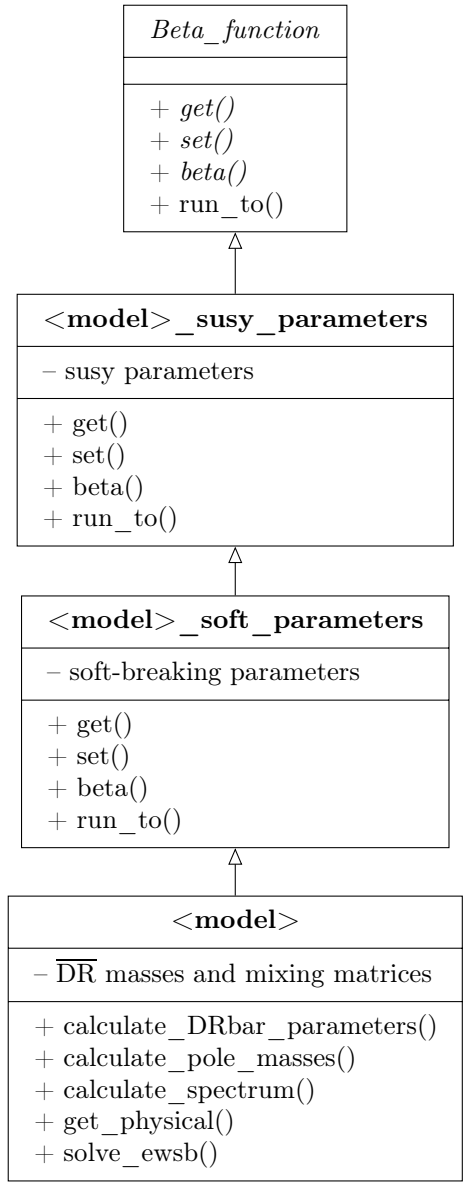
\begin{figure}
  \centering
  \tikzumlset{fill class=white}
  \begin{tikzpicture}
    \umlclass[x=0, y=8, type=abstract]{Beta\_function}{
    }{
      + \umlvirt{get()}\\
      + \umlvirt{set()}\\
      + \umlvirt{beta()}\\
      + run\_to()
    }
    \umlclass[x=0, y=4]{<model>\_susy\_parameters}{
      -- susy parameters
    }{
      + get()\\
      + set()\\
      + beta()\\
      + run\_to()
    }
    \umlclass[x=0, y=0]{<model>\_soft\_parameters}{
      -- soft-breaking parameters
    }{
      + get()\\
      + set()\\
      + beta()\\
      + run\_to()
    }
    \umlclass[x=0, y=-4.2]{<model>}{
      -- \DRbar\ masses and mixing matrices
    }{
      + calculate\_DRbar\_parameters()\\
      + calculate\_pole\_masses()\\
      + calculate\_spectrum()\\
      + get\_physical()\\
      + solve\_ewsb()
    }
    \umlinherit{<model>\_susy\_parameters}{Beta\_function}
    \umlinherit{<model>\_soft\_parameters}{<model>\_susy\_parameters}
    \umlinherit{<model>}{<model>\_soft\_parameters}
  \end{tikzpicture}
  \caption{Model class hierarchy.}
  \label{fig:parameter-classes}
\end{figure}

The parameters of the model together with their RGEs, mass
matrices, self-energies and EWSB equations are stored at the C++ level
in the model class hierarchy, which is shown in the UML diagram in
\figref{fig:parameter-classes}.

The top of the hierarchy is formed by the \code{Beta_function}
interface class, which defines the basic RGE running interface.  It
provides the interface function \code{run_to()}, which integrates the
RGEs up to a given scale using an adaptive Runge-Kutta algorithm.
This algorithm uses the pure virtual functions \code{get()},
\code{set()} and \code{beta()}, which need to be implemented by a
derived class.  The \code{get()} and \code{set()} functions return and
set the model parameters in form of a vector, respectively.  The
\code{beta()} method returns the $\beta$-function for each parameter
in form of a vector as well.

All model parameters and their $\beta$-functions are contained in
the first and second derived classes.  The structure of the
$\beta$-functions of a general supersymmetric model
\cite{Jones:1974pg,Jones:1983vk,West:1984dg,Martin:1993yx,Yamada:1993ga,Jack:1994kd,Jack:1994rk,Yam94,MV94,Fonseca:2011vn,Goodsell:2012fm,Sperling:2013eva,Sperling:2013xqa}
allows to split these parameters into two classes:
\begin{enumerate}
\item \emph{SUSY parameters:} gauge couplings, superpotential
  parameters and VEVs and
\item \emph{soft-breaking parameters} \cite{Girardello:1981wz}: soft
  linear scalar terms, soft bilinear scalar interactions, soft
  trilinear scalar interactions, soft quadlinear scalar interactions,
  soft gaugino mass terms and soft
  scalar squared masses.
\end{enumerate}
The $\beta$-functions of the SUSY parameters in general depend only on
the SUSY parameters and are independent of the soft-breaking
parameters.  However, the $\beta$-functions of the soft-breaking
parameters depend on all model parameters in general.
This property is reflected in the C++ code: The class
\code{<model>_susy_parameters} directly inherits from
\code{Beta_functions} and implements the $\beta$-functions of the SUSY
parameters.  The class of soft-breaking parameters
\code{<model>_soft_parameters} in turn inherits from
\code{<model>_susy_parameters} and implements the $\beta$-functions of
the soft-breaking parameters in terms of all model parameters.  The so
constructed class hierarchy allows to (i) use the RGE running of all
model parameters via the common \code{Beta_function} interface and to
(ii) run the SUSY parameters independently of the soft-breaking
parameters.

\fs creates these two classes from the model parameters defined in the
SARAH model file.  The corresponding one- and two-loop
$\beta$-functions are calculated algebraically using SARAH's
\code{CalcRGEs[]} routine, converted to C++ form and written into the
corresponding \code{beta()} functions.  These two classes then allow
to use renormalization group running of all model parameters.

At the bottom of the hierarchy stands the actual model class, which
uses the \DRbar\ parameters from the parent classes to calculate
\DRbar\ and pole mass spectra.  These two calculations are performed
in the \code{calculate_DRbar_masses()} and
\code{calculate_pole_masses()} functions, which make use of the mass
matrices and self-energies obtained from SARAH.  The calculation of
the pole mass spectrum will be explained in detail in
\secref{sec:PoleMasses}.  The resulting masses can be obtained by
calling \code{get_physical()}.  The \code{calculate_spectrum()}
function combines these two spectrum computations into one call.  In
addition, the model class provides a \code{solve_ewsb()} method, which
solves the electroweak symmetry breaking equations numerically at the
loop level.  This function is explained in the next section.

\subsection{Boundary conditions}
\label{sec:boundary-conditions}

As described in \secref{sec:modfile}, the user defines three boundary
conditions in the \fs model file at the \mathematica level.  These
boundary conditions are converted to C++ form and are put into
classes, which implement the common \code{Constraint<Two\_scale>}
interface.  This interface has the form:
\begin{lstlisting}[language=C++]
template<>
class Constraint<Two_scale> {
public:
   virtual ~Constraint() {}
   virtual void apply() = 0;
   virtual double get_scale() const = 0;
};
\end{lstlisting}
The \code{get_scale()} function is supposed to return the
renormalization scale at which the constraint is to be imposed.  The
\code{apply()} method imposes the constraint by setting model
parameters to values as chosen by the user.  The three boundary
condition classes are generated as follows:
\begin{itemize}
\item The \emph{high-scale constraint} is intended to set boundary
  conditions on the model parameters at some very high scale, e.g.\
  the GUT scale $M_X$.  The high-scale is defined by the value given
  in the variable \code{HighScale}.  In the CMSSM example model file
  in \secref{sec:modfile} it is defined to be the unification scale
  $M_X$ where $g_1(M_X) = g_2(M_X)$.  The \code{apply()} function is
  implemented by setting model parameters to the values defined in the
  \code{HighScaleInput} variable.

\item The \emph{SUSY-scale constraint} is intended to set boundary
  conditions at the mass scale $M_S$ of the SUSY particles.  The value
  of $M_S$ is defined in the model file variable \code{SUSYScale}.  In
  the example model file in \secref{sec:modfile} it is given by the
  expression written in Eq.~\eqref{eq:definition_of_MS}.  The
  \code{apply()} function for this constraint sets the model
  parameters to the values defined in \code{SUSYScaleInput}.
  Afterwards, \code{apply()} solves the EWSB equations at the loop
  level by adjusting the parameters given in
  \code{EWSBOutputParameters} such that the effective Higgs potential
  is minimized.  See \secref{sec:ewsb} for a more detailed description
  of the EWSB in \fs.

\item The \emph{low-scale constraint} is intended to match the SUSY
  model to the Standard Model at the scale $M_Z$.  It does so by
  calculating the gauge couplings of the SUSY model from the known
  Standard Model quantities
  $\alpha_{\text{e.m.},\text{SM}}^{(5),\text{\MSbar}}(M_Z)$,
  $\alpha_{\text{s},\text{SM}}^{(5),\text{\MSbar}}(M_Z)$, $M_Z$ and
  $M_W$.  This calculation is explained in
  \secref{sec:calculation-of-gauge-couplings}.  Optionally, the Yukawa
  couplings of the SUSY model can be calculated automatically from the
  Standard Model fermion masses.  See
  \secref{sec:calculation-of-yukawa-couplings} for more details.  In
  addition to the gauge and Yukawa couplings, the model parameter
  constraints given in \code{LowScaleInput} are imposed here.
\end{itemize}

\subsubsection{Calculation of the gauge couplings $g_i(M_Z)$}
\label{sec:calculation-of-gauge-couplings}

The low-scale constraint matches the SUSY model to the Standard Model.
Currently \fs allows only SUSY models with semisimple gauge groups,
which contain the Standard Model gauge group $SU(3)\times SU(2)\times
U(1)$ as factor.  This enables \fs to directly identify the strong,
left-handed and hypercharge gauge couplings $g_i$ ($i=1,2,3$).

The low-scale constraint automatically calculates the \DRbar\ gauge
couplings $g_{i,\text{susy}}^{\text{\DRbar}}(M_Z)$ in the SUSY model
at the scale $M_Z$.  It starts from the known electromagnetic and
strong \MSbar\ couplings in the Standard Model including only $5$
quark flavours
$\alpha_{\text{e.m.},\text{SM}}^{(5),\text{\MSbar}}(M_Z) = 1/127.944$
and $\alpha_{\text{s},\text{SM}}^{(5),\text{\MSbar}}(M_Z) = 0.1185$
\cite{Beringer:1900zz}.  These are converted to the electromagnetic
and strong \DRbar\ couplings in the SUSY model
$e_{\text{susy}}^{\text{\DRbar}}(M_Z)$ and
$g_{3,\text{susy}}^{\text{\DRbar}}(M_Z)$ as
\begin{align}
  \alpha_{\text{e.m.},\text{susy}}^{\text{\DRbar}}(M_Z) &=
  \frac{\alpha_{\text{e.m.},\text{SM}}^{(5),\text{\MSbar}}(M_Z)}{1 -
    \Delta\alpha_{\text{e.m.},\text{SM}}(M_Z) -
    \Delta\alpha_{\text{e.m.},\text{susy}}(M_Z)} ,\\
    e_{\text{susy}}^{\text{\DRbar}}(M_Z) &=
    \sqrt{4\pi\alpha_{\text{e.m.},\text{susy}}^{\text{\DRbar}}(M_Z)}, \\
  \alpha_{\text{s},\text{susy}}^{\text{\DRbar}}(M_Z) &=
  \frac{\alpha_{\text{s},\text{SM}}^{(5),\text{\MSbar}}(M_Z)}{1 -
    \Delta\alpha_{\text{s},\text{SM}}(M_Z)
    - \Delta\alpha_{\text{s},\text{susy}}(M_Z)} ,\\
  g_{3,\text{susy}}^{\text{\DRbar}}(M_Z) &=
  \sqrt{4\pi\alpha_{\text{s},\text{susy}}^{\text{\DRbar}}(M_Z)} .
\end{align}
The $\Delta\alpha_i(\mu)$ are threshold corrections and read
\begin{align}
  \Delta\alpha_{\text{e.m.},\text{SM}}(\mu) &=
  \frac{\alpha_\text{e.m.}}{2\pi} \left[\frac{1}{3}
    - \frac{16}{9} \log{\frac{m_t}{\mu}} \right],\\
  \Delta\alpha_{\text{e.m.},\text{susy}}(\mu) &=
  \frac{\alpha_\text{e.m.}}{2\pi} \left[ -\sum_{\text{susy particle }
      i}
    F_i T_i \log{\frac{m_i}{\mu}} \right],\\
  \Delta\alpha_{\text{s},\text{SM}}(\mu) &=
  \frac{\alpha_\text{s}}{2\pi} \left[
    -\frac{2}{3} \log{\frac{m_t}{\mu}} \right],\\
  \Delta\alpha_{\text{s},\text{susy}}(\mu) &=
  \frac{\alpha_\text{s}}{2\pi}\left[ \frac{1}{2}-\sum_{\text{susy
        particle } i} F_i T_i \log{\frac{m_i}{\mu}} \right] ,
\end{align}
where the sums on the right-hand sides run over all electrically and
color charged fields absent from the Standard Model.  The constants
$T_i$ are the Dynkin indices of the representation of particle~$i$
with respect to the gauge group, and $F_i$ are particle-type specific
constants \cite{Hall:1980kf}
\begin{align}
  F_i =
  \begin{cases}
    2/3 & \text{if particle $i$ is a Majorana fermion},\\
    4/3 & \text{if particle $i$ is a Dirac fermion},\\
    1/6 & \text{if particle $i$ is a real scalar},\\
    1/3 & \text{if particle $i$ is a complex scalar}.
  \end{cases}
\end{align}
Afterwards, the user-defined expression for the Weinberg angle
$\theta_W$ in terms of $M_{W,\text{susy}}^{\text{\DRbar}}(M_Z)$ and
$M_{Z,\text{susy}}^{\text{\DRbar}}(M_Z)$ (defined by the user in the
SARAH model file) is used to calculate $\theta_W$ in the SUSY model in
the \DRbar\ scheme.  In the MSSM, for example, it yields
\begin{align}
  \theta_{W,\text{susy}}^{\text{\DRbar}}(M_Z) &= \arcsin\sqrt{1
    -
    \left(\frac{M_{W,\text{susy}}^{\text{\DRbar}}(M_Z)}{M_{Z,\text{susy}}^{\text{\DRbar}}(M_Z)}\right)^2}
  .
\end{align}
In a model with a Higgs triplet the relation looks like
\begin{align}
  \theta_{W,\text{susy}}^{\text{\DRbar}}(M_Z) &= \arcsin\sqrt{1 -
    \frac{\left(M_{W,\text{susy}}^{\text{\DRbar}}(M_Z)\right)^2 -
      g_2^2v_T^2}{\left(M_{Z,\text{susy}}^{\text{\DRbar}}(M_Z)\right)^2}},
\end{align}
where $v_T$ is the vacuum expectation value of the scalar Higgs
triplet field.  In models with an additional $U(1)'$ gauge group, the
additional $Z'$ gauge boson can mix with the $Z$ boson.  In such
models the Weinberg angle can be defined by parametrizing the
$Z$--$Z'$ mixing matrix as
\begin{align}
  \begin{pmatrix}
    \cos\theta_W & -\sin\theta_W \cos\theta_W' &  \sin\theta_W \sin\theta_W' \\
    \sin\theta_W &  \cos\theta_W \cos\theta_W' & -\cos\theta_W \sin\theta_W' \\
    0 & \sin\theta_W' & \cos\theta_W'
  \end{pmatrix},
\end{align}
where $\theta_W'$ is the $Z$--$Z'$ mixing angle.
The running \DRbar\ $W$ and $Z$ boson masses are calculated in each
iteration from the corresponding pole masses as
\begin{align}
  \left(M_{W,\text{susy}}^{\text{\DRbar}}(M_Z)\right)^2 &=
  M_W^2 + \re \Pi_{WW}^T(p^2 = M_W^2, \mu=M_Z) ,\\
  \left(M_{Z,\text{susy}}^{\text{\DRbar}}(M_Z)\right)^2 &=
  M_Z^2 + \re \Pi_{ZZ}^T(p^2 = M_Z^2, \mu=M_Z) ,
\end{align}
where $M_W = 80.404\unit{GeV}$ and $M_Z = 91.1876\unit{GeV}$
\cite{Beringer:1900zz}.  Having $e_{\text{susy}}^{\text{\DRbar}}(M_Z)$
and $\theta_{W,\text{susy}}^{\text{\DRbar}}(M_Z)$ allows to calculate
the (GUT-normalized) $U(1)_Y$ and $SU(2)_L$ gauge couplings in the
SUSY model.  In the MSSM they read for instance
\begin{align}
  g_{1,\text{susy}}^{\text{\DRbar}}(M_Z) &=
  \sqrt{\frac{5}{3}} \frac{e_{\text{susy}}^{\text{\DRbar}}(M_Z)}{\cos\theta_{W,\text{susy}}^{\text{\DRbar}}(M_Z)} ,\\
  g_{2,\text{susy}}^{\text{\DRbar}}(M_Z) &=
  \frac{e_{\text{susy}}^{\text{\DRbar}}(M_Z)}{\sin\theta_{W,\text{susy}}^{\text{\DRbar}}(M_Z)} .
\end{align}

\subsubsection{Calculation of the Yukawa couplings $y_f(M_Z)$}
\label{sec:calculation-of-yukawa-couplings}

The considered SUSY model is required to contain the three generations
of Standard Model quarks and leptons.  If these particles acquire
their masses due to Yukawa interactions with Higgs doublets, then the
$3\times 3$ Yukawa matrices $y_f$ ($f=u,d,e$) can be calculated
automatically in the \DRbar\ scheme at the scale $M_Z$ from the known
Standard Model fermion masses by setting $y_f$ to the value
\code{Automatic} in the \fs model file.
This is done for example in
the CMSSM model file in \secref{sec:modfile}.  In this case \fs
expresses the Yukawa couplings in terms of the fermion mass matrices
$m_u$, $m_d$, $m_e$.  In the MSSM, for example, these relations read
in the SLHA convention \cite{Allanach:2008qq}
\begin{align}
  y_u^{\text{\DRbar}}(M_Z) &= \frac{\sqrt{2} m_{u}^T}{v_u} , &
  y_d^{\text{\DRbar}}(M_Z) &= \frac{\sqrt{2} m_{d}^T}{v_d} , &
  y_e^{\text{\DRbar}}(M_Z) &= \frac{\sqrt{2} m_{e}^T}{v_d} ,
\end{align}
where the superscript $T$ means transposition of a matrix.  The
fermion mass matrices are composed as
\begin{align}
  m_u = \diag(m_{u}^{\userinput}, m_{c}^{\userinput}, m_{t,\text{susy}}^{\text{\DRbar}}(M_Z)) ,\\
  m_d = \diag(m_{d}^{\userinput}, m_{s}^{\userinput}, m_{b,\text{susy}}^{\text{\DRbar}}(M_Z)) ,\\
  m_e = \diag(m_{e}^{\userinput}, m_{\mu}^{\userinput}, m_{\tau,\text{susy}}^{\text{\DRbar}}(M_Z)),
\end{align}
where the values for $m_{u,c,d,s,e,\mu}^{\userinput}$ are read from
the \code{SMINPUTS} block of the SLHA input file \cite{Skands:2003cj}.
The CKM mixing matrix is currently set to unity and $CP$-violating
phases are set to zero.  The third generation quark masses are
calculated in the \DRbar scheme from the SLHA user input quantities
$m_t^\text{pole}$, $m_{b,\text{SM}}^{\text{\MSbar}}(M_Z)$ and
$m_{\tau,\text{SM}}^{\text{\MSbar}}(M_Z)$ \cite{Skands:2003cj}.  In
detail, the top quark \DRbar mass is calculated as
\begin{align}
  \begin{split}
    m_{t,\text{susy}}^{\text{\DRbar}}(\mu) &= m_t^\text{pole} +
    \re\Sigma_{t}^{S}(m_t^\text{pole}) \\
    &\phantom{=\;} + m_t^\text{pole}
    \left[ \re\Sigma_{t}^{L}(m_t^\text{pole}) +
      \re\Sigma_{t}^{R}(m_t^\text{pole}) + \Delta
      m_t^{(1),\text{qcd}} + \Delta m_t^{(2),\text{qcd}} \right] ,
  \end{split}
\end{align}
where the $\Sigma_{t}$ is the top one-loop self-energy without QCD
contributions.  The labels $L,R,S$ denote the left-, right- and
non-polarized part of the self-energy, $\Sigma_{t}$.  The separated QCD corrections
$\Delta m_t^{(1),\text{qcd}}$ and $\Delta m_t^{(2),\text{qcd}}$ are
taken from \cite{Avdeev:1997sz,Bednyakov:2002sf} and read
\begin{align}
  \Delta m_t^{(1),\text{qcd}} &= -\frac{g_3^2}{12 \pi^2} \left[5-3 \log\left(\frac{m_t^2}{\mu^2}\right)\right],
  \label{eq:top-selfenergy-qcd-1L}\\
  \begin{split}
    \Delta m_t^{(2),\text{qcd}} &= \left(\Delta
      m_t^{(1),\text{qcd}}\right)^2 \\
    &\phantom{=\;} - \frac{g_3^4}{4608 \pi^4} \Bigg[396
    \log^2\left(\frac{m_t^2}{\mu^2}\right)-1476
    \log\left(\frac{m_t^2}{\mu^2}\right)
    -48 \zeta(3)+2011+16 \pi ^2 (1+\log 4)\Bigg].
  \end{split}
  \label{eq:top-selfenergy-qcd-2L}
\end{align}
In Eqs.\ \eqref{eq:top-selfenergy-qcd-1L} and
\eqref{eq:top-selfenergy-qcd-2L} $m_t$ denotes the \DRbar mass of the
top quark.
The \DRbar mass of the bottom quark is calculated as
\cite{Avdeev:1997sz,Baer:2002ek,Skands:2003cj}
\begin{align}
  m_{b,\text{susy}}^{\text{\DRbar}}(\mu) &=
  \frac{m_{b,\text{SM}}^{\text{\DRbar}}(\mu)}{1 -
    \re\Sigma_{b}^{S,\text{heavy}}(m_{b,\text{SM}}^\text{\MSbar})/m_b
    - \re\Sigma_{b}^{L,\text{heavy}}(m_{b,\text{SM}}^\text{\MSbar}) -
    \re\Sigma_{b}^{R,\text{heavy}}(m_{b,\text{SM}}^\text{\MSbar})} ,\\
  m_{b,\text{SM}}^{\text{\DRbar}}(\mu) &=
  m_{b,\text{SM}}^{\text{\MSbar}}(\mu) \left(1 - \frac{\alpha_s}{3
      \pi} - \frac{23}{72} \frac{\alpha_s^2}{\pi^2} + \frac{3
      g_2^2}{128 \pi^2} + \frac{13 g_Y^2}{1152 \pi^2}\right) ,
\end{align}
where $\tan\beta$ enhanced loop self-energy corrections are resummed.
Finally, the \DRbar mass of the $\tau$ is calculated as
\begin{align}
  \begin{split}
    m_{\tau,\text{susy}}^{\text{\DRbar}}(\mu) &=
    m_{\tau,\text{SM}}^{\text{\DRbar}}(\mu) +
    \re\Sigma_{\tau}^{S,\text{heavy}}(m_{\tau,\text{SM}}^\text{\MSbar}) \\
    &\phantom{=\;} + m_{\tau,\text{SM}}^{\text{\DRbar}}(\mu) \left[
      \re\Sigma_{\tau}^{L,\text{heavy}}(m_{\tau,\text{SM}}^\text{\MSbar})
      +
      \re\Sigma_{\tau}^{R,\text{heavy}}(m_{\tau,\text{SM}}^\text{\MSbar})
    \right] ,
  \end{split}\\
  m_{\tau,\text{SM}}^{\text{\DRbar}}(\mu) &= m_{\tau,\text{SM}}^{\text{\MSbar}}(\mu)
  \left(1 - 3 \frac{g_Y^2 - g_2^2}{128 \pi^2}\right).
\end{align}
In the above equations $\Sigma_{b,\tau}^{\text{heavy}}$ are the
one-loop self-energies of the bottom and $\tau$, where contributions
from the gluon and photon are omitted.  To convert the fermion masses
from the \MSbar to the \DRbar scheme the Yukawa coupling conversion
from \cite{Martin:1993yx} is used and it is assumed that the VEV is
defined in the \DRbar scheme.

\subsubsection{Electroweak symmetry breaking}
\label{sec:ewsb}

\fs assumes that each SUSY model contains Higgs bosons, which trigger
a spontaneous breaking of the electroweak symmetry.  The corresponding
EWSB consistency conditions are formulated in \fs at the one-loop
level as
\begin{align}
  0 &= \frac{\partial V^\text{tree}}{\partial v_i} - t_i,
  & & (i=1,\ldots,N)
  \label{eq:one-loop-ewsb-eq}
\end{align}
where $V^\text{tree}$ is the tree-level Higgs potential, $v_i$ is the
VEV corresponding to the Higgs field $H_i$ and $t_i$ is the one-loop
tadpole diagram of $H_i$.  Already at the tree-level ($t_i = 0$) Eqs.\
\eqref{eq:one-loop-ewsb-eq} can have multiple solutions, depending on
which parameters are chosen to be fixed by these equations.  A
well-known example is the real MSSM, where the $\mu$-parameter is
chosen to be fixed by the EWSB equations.  In this model, the EWSB
equations, Eqs.\ \eqref{eq:one-loop-ewsb-eq} can only determine
$|\mu|$, while the sign of $\mu$ is not fixed.  This results in two
solutions for $\mu$ of the form
\begin{align}
  \mu = \sign\mu \cdot |\mu| .
  \label{eq:EWSB-solution-for-mu}
\end{align}
Currently, \fs handles only real model parameters and renormalizable
theories, which restricts the number of possible multiple solutions of
Eqs.\ \eqref{eq:one-loop-ewsb-eq} to be less than or equal to $4N$.  In
some cases, as for example in the real MSSM, the different solutions
are related by one or more global signs for the parameters, as for example in
Eq.~\eqref{eq:EWSB-solution-for-mu}.  \fs recognizes such cases and
introduces these global signs as additional free parameters to allow
the user to choose between the different solutions.  The case of the
real MSSM Eq.~\eqref{eq:EWSB-solution-for-mu} is therefore handled
automatically in \fs, because $\sign\mu$ is automatically introduced
as an additional free parameter.

If, however, the different solutions are not related by global signs,
then \fs writes all solutions to the file
\code{models/<model>/<model>_tree_level_EWSB_solution.m}.  The user
has then the option to pick a particular solution by setting it in the
\code{TreeLevelEWSBSolution} variable in the model file.  This is for
example the case in the $Z_3$-violating NMSSM (\code{SMSSM}), which is shipped with \fs, where
$\mu$ is chosen to be fixed by the EWSB equations: There the
tree-level solution for $\mu$ has the form
\begin{align}
  \mu = -\frac{v_s \lambda}{\sqrt{2}} + \sign X \cdot \sqrt{\ldots}
\end{align}
with $\sign X$ as free parameter.  When \fs solves the tree-level EWSB
equations for $\mu$, $B\mu$ and $\xi_S$ it finds the two solutions and
writes them to \code{models/SMSSM/SMSSM_tree_level_EWSB_solution.m} in
the form
\begin{lstlisting}[language=Mathematica]
{
   {{B[\[Mu]] -> ...}},
   {
      {\[Mu] -> (-20*Sqrt[2]*vd*vS*(\[Lambda] + conj[\[Lambda]])
                 - Sqrt[...])/(80*vd)},
      {\[Mu] -> (-20*Sqrt[2]*vd*vS*(\[Lambda] + conj[\[Lambda]])
                 + Sqrt[...])/(80*vd)}
   },
   {{L[L1] -> ...}}
}
\end{lstlisting}
In the above solutions the dots stand for the full expression, which
is left out here for better readability.  Inspecting the two above
solutions for $\mu$, one finds that both can be parametrized by an
additional free sign in front of the \code{Sqrt[...]}.  The user can
now introduce an additional free sign by hand in the \code{MINPAR}
block
\begin{lstlisting}[language=Mathematica]
MINPAR = {
   {1, m0},
   {2, m12},
   {3, TanBeta},
   {4, Sign[X]},  (* <-- additional free sign *)
   {5, Azero}
};
\end{lstlisting}
and set the the tree-level solution, parametrized in terms of
\code{Sign[X]}, in the \code{TreeLevelEWSBSolution} variable in
the model file:
\begin{lstlisting}
TreeLevelEWSBSolution = {
   { B[\[Mu]], ... },
   { \[Mu], (-20*Sqrt[2]*vd*vS*(\[Lambda] + conj[\[Lambda]])
             + Sign[X] * Sqrt[...])/(80*vd) },
   { L[L1], ... }
};
\end{lstlisting}
One can now choose between the two solutions by setting entry $4$ in
the \code{MINPAR} block of the SLHA input file to either $+1$ or $-1$.
See \code{model_files/SMSSM/FlexibleSUSY.m.in} for a complete example
model file.  If the user decides to not pick a particular solution by
leaving the variable \code{TreeLevelEWSBSolution} empty, \fs tries to
find a solution to the tree-level EWSB equations numerically via an
iteration.  In this case, however, the user does not have the option
to choose between the different solutions.

If a solution of the tree-level EWSB equations has been found, the
one-loop equations \eqref{eq:one-loop-ewsb-eq} are solved
simultaneously using the iterative multi-dimensional root finder
algorithm \code{gsl_multiroot_fsolver_hybrid} from the GNU Scientific
Library (GSL).  If no root can be found, the
\code{gsl_multiroot_fsolver_hybrids} algorithm is tried as
alternative, which uses a variable step size but might be a little
slower.

In the CMSSM example from \secref{sec:modfile} the
Eqs.~\eqref{eq:one-loop-ewsb-eq} are expressed in the form of the
following C++ function:
\begin{lstlisting}[language=C++]
int MSSM<Two_scale>::tadpole_equations(const gsl_vector* x, void* params,
                                       gsl_vector* f)
{
   ...

   double tadpole[number_of_ewsb_equations];

   model->set_BMu(gsl_vector_get(x, 0));
   model->set_Mu(INPUT(SignMu) * Abs(gsl_vector_get(x, 1)));

   // calculate tree-level tadpole eqs.
   tadpole[0] = model->get_ewsb_eq_vd();
   tadpole[1] = model->get_ewsb_eq_vu();

   // subtract one-loop tadpoles
   if (ewsb_loop_order > 0) {
      model->calculate_DRbar_masses();
      tadpole[0] -= Re(model->tadpole_hh(0));
      tadpole[1] -= Re(model->tadpole_hh(1));
   }

   for (std::size_t i = 0; i < number_of_ewsb_equations; ++i)
      gsl_vector_set(f, i, tadpole[i]);

   return GSL_SUCCESS;
}
\end{lstlisting}
The function parameter \code{x} is the vector of EWSB output
parameters (defined in \code{EWSBOutputParameters}) and \code{f} is a
vector which contains the one-loop EWSB Eqs.\
\eqref{eq:one-loop-ewsb-eq}.  This \code{tadpole_equations()} function
is passed to the root finder, which searches for values of the model
parameters $\mu$ and $B\mu$ until the Eqs.\
\eqref{eq:one-loop-ewsb-eq} are fulfilled.

If higher accuracy is required additional routines with higher order
corrections can be added by setting \code{UseHiggs2LoopMSSM = True}
in the model file.  For example in the MSSM by default \fs adds
two-loop Higgs FORTRAN routines supplied by P.~Slavich from
\cite{Dedes:2002dy,Dedes:2003km}
to add two-loop corrections of $\oatas$, $\oabas$, $\oatq$,
$\oabq$, $\oatauq$ and $\oatab$.  In the NMSSM the same contributions
can be added by setting \code{UseHiggs2LoopNMSSM = True} in the model
file.

\subsection{Tree-level spectrum}
\label{sec:TreeLevelSpectrum}
The tree-level \DRbar masses are calculated from the \DRbar\ model
parameters by diagonalizing the mass matrices returned from
\code{SARAH`MassMatrix[]}.  The numerical singular value decomposition
is performed by the Eigen library routine \code{Eigen::JacobiSVD} for
matrices with less than four rows and columns, and the LAPACK routines
\code{zgesvd}, \code{dgesvd} for larger matrices.  For the other types
of diagonalization, \code{Eigen::SelfAdjointEigenSolver} from Eigen is
used regardless of the matrix size.  Note, that \fs uses double
precision floating point data types with $15$ significant digits to
store the mass matrices and the mass eigenvalues.  In case a particle
multiplet contains a very split mass hierarchy, where the mass
difference between the smallest and the largest mass in the multiplet
is of the order or greater than 10 orders of magnitude, double
precision data types are no longer sufficient.  In this case we
recommend to either split the multiplet into sub-multiplets with
smaller mass hierarchies, or integrate out the heavy states.

\fs uses the following conventions for the diagonalization: A mass
matrix $M^2$ for real scalar fields $\phi_i$ is diagonalized with an
orthogonal matrix $O$ as
\begin{align}
  \Lagr_{m,\text{real scalar}}
  &= - \frac{1}{2} \phi^T M^2 \phi
  = - \frac{1}{2} (\phi^m)^T M^2_D \phi^m, \\
  \qquad M^2 &= (M^2)^T ,
  \qquad \phi^m = O \phi ,
  \qquad M^2_D = O M^2 O^T ,
  \qquad O^T O = \unity ,
\end{align}
where $M^2_D$ is diagonal and $\phi^m_i$ are the mass eigenstates.  In
case of complex scalar fields $\phi_i$ we use
\begin{align}
  \Lagr_{m,\text{complex scalar}}
  &= - \phi^\dagger M^2 \phi
  = - (\phi^m)^\dagger M^2_D \phi^m, \\
  \qquad M^2 &= (M^2)^\dagger ,
  \qquad \phi^m = U \phi ,
  \qquad M^2_D = U M^2 U^\dagger ,
  \qquad U^\dagger U = \unity .
\end{align}
A (possibly complex) symmetric mass matrix $Y$ for Weyl spinors
$\psi_i$ is diagonalized as
\begin{align}
  \Lagr_{m,\text{fermion}}^\text{symm.}
  &= - \frac{1}{2} \psi^T Y \psi + \text{h.c.}
  = - \frac{1}{2} \chi^T Y_D \chi + \text{h.c.}, \\
  \qquad Y &= Y^T ,
  \qquad Y_D = Z^* Y Z^\dagger ,
  \qquad \chi = Z \psi ,
  \qquad Z^\dagger Z = \unity ,
\end{align}
where $Y_D$ is diagonal and $\chi_i$ are the mass eigenstates.  The
phases of $Z$ are chosen such that all mass eigenvalues are positive.
In case of a non-symmetric mass matrix $X$ for Weyl spinors $\psi_i$
we use
\begin{align}
  \Lagr_{m,\text{fermion}}^\text{svd}
  &= - (\psi^-)^T X \psi^+ + \text{h.c.}
  = - (\chi^-)^T X_D \chi^+ + \text{h.c.}, \\
  \qquad \chi^+ &= V \psi^+ ,
  \qquad \chi^- = U \psi^- ,
  \qquad X_D = U^* X V^{-1} ,
  \qquad U^\dagger U = \unity = V^\dagger V ,
\end{align}
where we are again choosing the phases of $U$ and $V$ such that all mass
eigenvalues are positive.

\subsection{Two-scale fixed point iteration}
\label{sec:TwoScaleFixedPointIteration}

As explained at the beginning of \secref{sec:SpecGenStruct}, the RGEs
plus the user-defined boundary conditions on the model parameters form
a boundary value problem.  \fs provides a default two-scale boundary
value problem solver, which tries to find a set of model parameters
consistent with all constraints at all scales.  It does so by running
iteratively between the scales of all boundary conditions, imposing
the constraints (by calling the corresponding \code{apply()} function)
and checking for convergence after each iteration.  This approach is
described in \cite{Barger:1993gh} originally for the MSSM and is widely
implemented in SUSY spectrum generators.
Despite sharing the same algorithm with others,
the boundary value problem solver class from \fs, named \code{RGFlow},
has two notable properties.
First, it extends the aforementioned procedure to towers of models.
If the problem involves more than one model,
\code{RGFlow} matches one model to the next after
running the model parameters to the matching scale.
Second, \code{RGFlow} is an abstract implementation of
the algorithm, unaware of physics, in that
it is free of hard-wired model-dependent code related to
RGEs, boundary or matching conditions, or initial guesses.
All these pieces of physics information are
carried by separate objects which one then
links to \code{RGFlow} to set up a boundary value problem.
This modular design makes it easy
to replace any of the above components,
as shall be demonstrated in \secref{sec:integrating-custom-built}.

\begin{figure}[tbh]
  \centering
  \begin{tikzpicture}[node distance = 2.2cm, auto]
    \tikzstyle{block} = [rectangle, draw, text width=16em, text centered, minimum height=3em]
    \tikzstyle{arrow} = [draw, -latex, thick]
    \node[block] (guess) {Guess $g_i(M_Z)$, $y_f(M_Z)$ and soft
      parameters at \code{LowScale}};
    \node[block,below of=guess] (MZ) {Calculate $g_i(M_Z)$, $y_f(M_Z)$ and apply
      low-scale boundary conditions (\code{LowScaleInput})};
    \path[arrow] (guess) -- node {run to \code{LowScale}} (MZ);
    \node[block,below of=MZ] (MX) {Apply high-scale boundary conditions
      (\code{HighScaleInput})};
    \path[arrow] (MZ) -- node {run to \code{HighScale}} (MX);
    \node[block,below of=MX] (MS) {Apply SUSY-scale boundary conditions
      (\code{SUSYScaleInput}) and solve EWSB};
    \path[arrow] (MX) -- node {run to \code{SUSYScale}} (MS);
    \path[-latex, thick] (MS.east) edge[bend right=90] node[right] {run to \code{LowScale}} (MZ.east);
    \node[block,below of=MS] (spec) {Calculate pole masses};
    \path[arrow,dashed] (MS) -- node[text width=16em] {if converged run to \code{SUSYScale}} (spec);
  \end{tikzpicture}
  \caption{Iterative two-scale algorithm to calculate the spectrum.}
  \label{fig:two-scale-algorithm}
\end{figure}
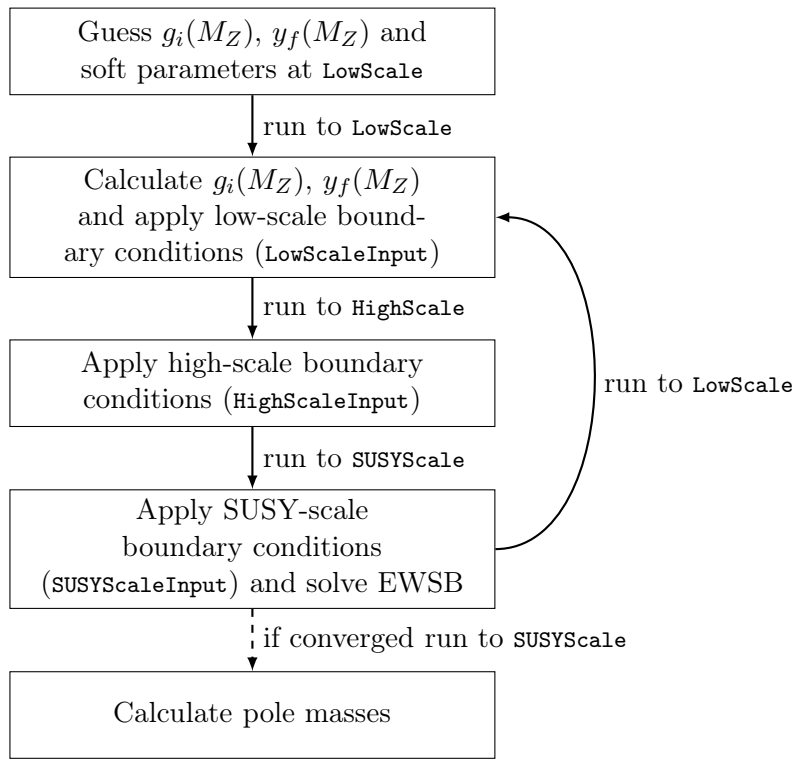

In more detail the two-scale algorithm used in \fs,
as applied to a problem with a single MSSM-like model,
works as follows, see also \figref{fig:two-scale-algorithm}:
\subparagraph{Initial guess:} The RG solver starts to guess all model
parameters at the low-scale.
\begin{enumerate}
\item At the $M_Z$ scale the gauge couplings $g_{1,2,3}$ are set to
  the known Standard Model values (ignoring threshold corrections).
\item The user-defined initial guess at the low-scale (defined in
  \code{InitialGuessAtLowScale}) is imposed.  In the example given in
  \secref{sec:modfile} the Higgs VEVs are set to
  \begin{align}
    v_d &= v \cos\beta, & v_u &= v \sin\beta ,
  \end{align}
  where $v=246.22\unit{GeV}$.  Afterwards, the Yukawa couplings
  $y_{u,d,e}$ of the SUSY model are set from the known Standard Model
  Yukawa couplings using the tree-level relations (ignoring SUSY
  radiative corrections).
\item The SUSY parameters are run to the user-supplied first guess of
  the high-scale (\code{HighScaleFirstGuess}).
\item The high-scale boundary condition is imposed (defined in
  \code{HighScaleInput}).  Afterwards, the user-defined initial guess
  for the remaining model parameters (defined in
  \code{InitialGuessAtHighScale}) is imposed.  In the example given in
  \secref{sec:modfile} the superpotential parameter $\mu$ is set to
  the value $1.0$ and its corresponding soft-breaking parameter $B\mu$
  is set to zero.
\item All model parameters are run to the first guess of the low-scale
  (\code{LowScaleFirstGuess}).
\item The EWSB eqs.\ are solved at the tree-level.
\item The \DRbar\ mass spectrum is calculated.
\end{enumerate}
At this point all model parameters are set to some initial values and
a first estimation of the \DRbar\ mass spectrum is known.  Now the
actual iteration starts
\subparagraph{Fixed-point iteration:}
\begin{enumerate}
\item \label{rge-step-one} All model parameters are run to the
  low-scale (\code{LowScale}).
  \begin{enumerate}
  \item The \DRbar\ mass spectrum is calculated.
  \item The low-scale is recalculated.  In the above example this step
    is trivial, because the low-scale is fixed to be $M_Z$.
  \item The \DRbar\ gauge couplings $g_{1,2,3}(M_Z)$ of the SUSY model
    are calculated using threshold corrections as described in
    \secref{sec:calculation-of-gauge-couplings}.
  \item The user-defined low-scale constraint is imposed
    (\code{LowScaleInput}).  In the example above, the Yukawa
    couplings are calculated automatically as described in
    \secref{sec:calculation-of-yukawa-couplings} and the Higgs VEVs
    are set to
    \begin{align}
      v_d(M_Z) &= \frac{2 M_Z^\text{\DRbar}(M_Z)}{\sqrt{0.6 g_1^2(M_Z) + g_2^2(M_Z)} \cos\beta(M_Z)}, \\
      v_u(M_Z) &= \frac{2 M_Z^\text{\DRbar}(M_Z)}{\sqrt{0.6 g_1^2(M_Z) + g_2^2(M_Z)} \sin\beta(M_Z)}.
    \end{align}
    Since the Hypercharge gauge coupling $g_1$ is GUT normalized, the
    normalization factor $\sqrt{3/5}$ has to be included in the above
    relations.
  \end{enumerate}
\item Run all model parameters to the high-scale (\code{HighScale}).
  \begin{enumerate}
  \item Recalculate the high-scale as
    \begin{align}
      M_X' = M_X \exp\left(\frac{g_2(M_X)-g_1(M_X)}{\beta_{g_1} -
          \beta_{g_2}}\right),
    \end{align}
    where $\beta_{g_i}$ is the two-loop $\beta$-function of the gauge
    coupling $g_i$.  The value $M_X'$ is used as new high-scale in the
    next iteration.
  \item Impose the high-scale constraint (\code{HighScaleInput}).  In
    the CMSSM example the following soft-breaking parameters are fixed
    to the universal values $m_0$, $M_{1/2}$ and $A_0$:
    \begin{align}
      A^f(M_X) &= A_0 & &(f=u,d,e),\\
      m_{H_i}^2(M_X) &= m_0^2 & &(i=1,2),\\
      m_{f}^2(M_X) &= m_0^2\mathbf{1} & &(f=q,l,d,u,e),\\
      M_{i}(M_X) &= M_{1/2} & &(i=1,2,3).
    \end{align}\label{eq:fs-cmssm-high-scale-bc}%
  \end{enumerate}
\item Run model parameters to the SUSY-scale (\code{SUSYScale}).
  \begin{enumerate}
  \item Calculate the \DRbar\ mass spectrum.
  \item Recalculate the SUSY-scale $M_S$ as
    \begin{align}
      M_S = \sqrt{\prod_{i=1}^6 m_{\tilde{u}_i}^{|(Z_u)_{i3}|^2 + |(Z_u)_{i6}|^2}} ,
    \end{align}
    where $m_{\tilde{u}_i}$ is the \DRbar\ mass of the $i$th up-type
    squark.
  \item Impose the SUSY-scale constraint (\code{SUSYScaleInput}).  In
    the example above, this step is trivial since
    \code{SUSYScaleInput} is set to be empty.
  \item Solve the EWSB equations iteratively at the loop level.  In
    the MSSM example from above leading two-loop corrections have been
    enabled by setting \code{UseHiggs2LoopMSSM = True}.  This will
    add two-loop tadpole contributions to the effective Higgs
    potential during the EWSB iteration.
  \end{enumerate}
\item If not converged yet, goto \ref{rge-step-one}.  Otherwise,
  finish the iteration.
\end{enumerate}
If the fixed-point iteration has converged, all \DRbar\ model
parameters are known at all scales between \code{LowScale} and
\code{HighScale}.  In this case all model parameters are run to the
SUSY-scale and the pole-mass spectrum is calculated as described in
\secref{sec:PoleMasses}.  If the user has chosen a specific output
scale for the running \DRbar\ model parameters by setting entry $12$
in block \code{MODSEL} in the SLHA input file, all model parameters
are finally run to the defined output scale.

During the fixed-point iteration several problems can appear.  First
of all, the iteration is not guaranteed to converge.  If the desired
accuracy goal is not achieved with the given maximum number of
iterations, \fs will set the \code{no_convergence} flag in the
\code{Problems} class.  This class monitors the problem status of the
spectrum generator during the iteration and can be obtained from the
model class via the \code{get_problems()} function.  Besides
non-convergence, solving the EWSB conditions
\eqref{eq:one-loop-ewsb-eq} numerically with the desired accuracy
might fail.  In this case the \code{no_ewsb} flag is set.
Furthermore, in intermediate iteration steps tachyonic states might
appear, which are ignored but nevertheless monitored in the
\code{Problems} class.  If tachyons still exist after the iteration
has converged the mass spectrum is marked as invalid by setting entry
$4$ in the \code{SPINFO} block in the SLHA output file.  Finally,
during the RG running some couplings might become non-perturbative.
In this case the iteration stops setting the \code{no_perturbative}
flag.

It is important to note that in the case of such problem points it is
non-trivial to judge whether this is because there is no physical
solution for the given parameter space point or a solution
exists but the fixed point iteration is unable to find the solution.
While \fs makes it as easy as possible to find spectra, when studying
new models a physical understanding of the model is still essential
and this can help the user determine why such problems arise.

Nonetheless \fs provides help for such cases in several ways. One may
adjust initial guesses specified in the \fs model file, such
as changing the choice of \code{HighScaleFirstGuess} or altering
\code{HighScaleMinimum} and \code{HighScaleMaximum} which can be used
to push the iteration back towards where the solution should be if it
gets off track. For experienced users the clear code structure also allows the
possibility of direct adaption of the code. 

Finally instead of tinkering with the two-scale solver one may wish to
replace it entirely. The modular design of \fs allows for the
solver for the boundary value problem to be replaced. An alternative
solver with potentially better convergence properties (at the expense of slower
speed) is already planned for a later release.

\subsection{Pole masses}
\label{sec:PoleMasses}
After the solver routine has finished and convergence has been
achieved, all \DRbar\ parameters consistent with the EWSB conditions,
low energy data and all user-supplied boundary conditions are
known at any scale between \code{LowScale} and \code{HighScale}.

The (physical) pole mass spectrum can now be calculated.  \fs uses the
full one-loop self-energies and tree-level mass matrices obtained from
SARAH to calculate the pole masses, which means finding the values $p$
that solve the equation
\begin{align}
  0 = \det\left[p^2\unity - m_{f,1L}(p^2)\right].
  \label{eq:pole-mass-def}
\end{align}
Here the one-loop mass matrix $m_{f,1L}(p^2)$ for field $f$ is given
in terms of the tree-level mass matrix $m_f$ and the self-energy
$\Sigma_f(p^2)$ as
\begin{align}
  &\text{scalars } \phi: &
  m_{\phi,1L}(p^2) &= m_{\phi} - \Sigma_\phi(p^2), \\
  &\text{Majorana fermions } \chi: &
  m_{\chi,1L}(p^2) &= m_{\chi} - \frac{1}{2}\Big[
    \Sigma_\chi^S(p^2) + \Sigma_\chi^{S,T}(p^2)
    + \Big( \Sigma_\chi^{L,T}(p^2) + \Sigma_\chi^R(p^2) \Big) m_{\chi} \notag \\
    &&&\phantom{= m_{\chi} - \frac{1}{2}\Big[}
    + m_{\chi} \Big( \Sigma_\chi^L(p^2) + \Sigma_\chi^{R,T}(p^2) \Big)
  \Big], \label{eq:majorana-loop-corrections} \\
  &\text{Dirac fermions } \psi: &
  m_{\psi,1L}(p^2) &= m_{\psi}
  - \Sigma_\psi^S(p^2)
  - \Sigma_\psi^R(p^2) m_{\psi}
  - m_{\psi} \Sigma_\psi^L(p^2) .
\end{align}
Eq.~\eqref{eq:pole-mass-def} can be solved by diagonalizing the
one-loop mass matrix $m_{f,1L}(p^2)$.  However, since $m_{f,1L}(p^2)$
depends on the momentum $p$, an iteration over $p$ must be performed.
Since this iteration can be very time consuming for large field
multiplets, \fs provides two approximative procedures with a shorter
run-time in addition to the iterative procedure.  In the \fs model
file the two approximative procedures are called \code{LowPrecision}
and \code{MediumPrecision}.  The iterative procedure is called
\code{HighPrecision}.  The procedure to be used can be set in the
model file for each field.  The default setting is
\begin{lstlisting}
DefaultPoleMassPrecision = MediumPrecision;
HighPoleMassPrecision    = {hh, Ah, Hpm};
MediumPoleMassPrecision  = {};
LowPoleMassPrecision     = {};
\end{lstlisting}
In the variable \code{DefaultPoleMassPrecision} the default pole mass
calculation precision to be used for all particles is selected.
Possible values are \code{LowPrecision}, \code{MediumPrecision} and
\code{HighPrecision}.  The values \code{LowPrecision} and
\code{MediumPrecision} correspond to the two approximations described
below and \code{HighPrecision} corresponds to the iterative
determination of the pole masses.  In the variables
\code{HighPoleMassPrecision}, \code{MediumPoleMassPrecision} and
\code{LowPoleMassPrecision} the pole mass calculation precision can be
changed for individual particles.  The settings in these variables
overwrite the setting in \code{DefaultPoleMassPrecision} for these
particles.  In the above example the pole masses of all particles are
calculated with \code{MediumPrecision}, except for the Higgs boson
pole masses, which are calculated with the iterative procedure
(\code{HighPrecision}).

The three different pole mass calculation procedures work as follows:
\begin{itemize}
\item \code{LowPrecision}: This option provides the lowest precision
  but is also the fastest one.  Here the one-loop mass matrix
  $m_{f,1L}^\text{low}$ is calculated exactly once as
  \begin{align}
    \forall i,j: (m_{f,1L}^\text{low})_{ij} = (m_{f,1L}(p^2 = m_{f_i}
    m_{f_j}))_{ij} ,
  \end{align}
  where $m_{f_i}$ is the $i$th mass eigenvalue of the tree-level mass
  matrix $m_f$.  Afterwards, $m_{f,1L}^\text{low}$ is diagonalized and
  the eigenvalues are interpreted as pole masses $m_{f_i}^\pole$.
  This method neglects terms of the form
  \begin{align}
    \left[(m_{f_k}^\pole)^2 - m_{f_i}m_{f_j}\right]
    \left.\frac{\partial m_{f,1L}(p^2)}{\partial
        p^2}\right|_{p^2=m_{f_i}m_{f_j}} ,
  \end{align}
  which are formally of two-loop order.  The method is imprecise if
  the self-energy corrections to the mass matrix are large or the
  tree-level mass spectrum of the multiplet is very split.  Note: We
  strongly discourage the use of this method for the determination of the
  Higgs pole masses, as the result will be very imprecise due to the
  large loop corrections.  \fs will print a warning if this method is
  used for any Higgs boson.

\item \code{MediumPrecision} (default): This option
  provides calculation with medium precision with a medium execution
  time.  Here the one-loop mass matrix $m_{f,1L}^\text{medium}$ is
  calculated $n$ times as
  \begin{align}
    (m_{f,1L}^\text{medium})_{ij}^{(k)} = (m_{f,1L}(p^2 =
    m_{f_k}^2))_{ij} , \qquad k = 1,\ldots,n ,
  \end{align}
  where $m_{f_k}$ is the $k$th mass eigenvalue of the tree-level mass
  matrix $m_f$.  Afterwards, each mass matrix
  $(m_{f,1L}^\text{medium})^{(k)}$ is diagonalized and the $k$th
  eigenvalue is interpreted as pole mass $m_{f_k}^\pole$.  Thereby for
  the $k$th eigenvalue two-loop terms of the form
  \begin{align}
    \left[(m_{f_k}^\pole)^2 - m_{f_k}^2\right] \left.\frac{\partial
        m_{f,1L}(p^2)}{\partial p^2}\right|_{p^2=m_{f_k}^2}
  \end{align}
  are neglected.  This method is imprecise if the self-energy
  corrections to the mass matrix are large.  Note, that this method is
  used in Softsusy to calculate the pole masses of the non-Higgs
  fields.

\item \code{HighPrecision}: This option solves
  Eq.~\eqref{eq:pole-mass-def} exactly by iterating over the momentum
  $p$.  It therefore provides the determination of the pole masses
  with highest precision, but has also the highest execution time.
  Here the one-loop mass matrix $m_{f,1L}^\text{high}$ is diagonalized
  $n$ times, as in the case of \code{MediumPrecision},
  resulting in $n$ pole masses $m_{f_k}^\pole$ ($k = 1,\ldots,n$).
  Afterwards, the diagonalization is repeated, this time using the
  calculated pole masses $m_{f_k}^\pole$ for the momentum calculation
  $p^2 = (m_{f_k}^\pole)^2$.  The iteration stops if convergence is
  reached.
\end{itemize}
A numerical comparison of the three different methods for a specific
CMSSM parameter point can be found in
\tabref{tab:pole-mass-precision-comparison}.  One finds that
(i) the calculated lightest $CP$-even Higgs pole mass, $m_h$, differs
about $1.2\unit{GeV}$ between \code{MediumPrecision} and
\code{HighPrecision}, due to large loop corrections.  Since the
experimental Higgs mass uncertainty is currently around
$0.4\unit{GeV}$ \cite{Beringer:1900zz}, we strongly recommend the use
of \code{HighPrecision} to calculate the Higgs pole mass (this is the
default).  Especially, if two-loop contributions to the Higgs tadpoles
and self-energies are added \code{MediumPrecision} must not be used,
because it neglects terms of two-loop order.  The Higgs boson mass for
the \code{LowPrecision} method is not given in the table, as it will
lead to an imprecise result and is therefore strongly discouraged to
be used, see \secref{sec:PoleMasses}.
(ii) The gluino pole mass, $m_{\tilde{g}}$, is given in the second
row.  Since the gluino does not mix with other particles, there is no
difference between \code{LowPrecision} and \code{MediumPrecision}.
Not neglecting the two-loop terms by using \code{HighPrecision}
increases the gluino mass about $0.5\unit{\%}$ for this parameter point.
(iii) The pole masses of the lightest and heaviest neutralinos,
$m_{\tilde{\chi}_1^0}$ and $m_{\tilde{\chi}_4^0}$, are given in the rows
$3$--$4$.  Since the
momentum-dependent loop-corrections to the lightest neutralino mass
are small for this parameter point, its pole mass varies only in the
sub-GeV range between the three methods.  However, the run-time of the
\code{LowPrecision} method is more than a factor $10$ smaller than of
the \code{HighPrecision}, due to the complicated structure of the loop
corrections in Eq.~\eqref{eq:majorana-loop-corrections}.
(iv) The pole masses of the lightest sfermions are given in rows
$5$--$7$.  Since these particles are contained in $6$-plets, the
run-time for the calculation of their pole masses is dramatically
increased by around a factor $20$ between \code{LowPrecision} and
\code{HighPrecision}.  However, since the change in the lightest
sfermion masses between the three different methods is less than
$\unit{0.3\%}$, one can consider calculating them with
\code{MediumPrecision} only.
\begin{table}[tbh]
  \centering
  \begin{tabular}{lrrr}
    \toprule
    & \code{LowPrecision}
    & \code{MediumPrecision}
    & \code{HighPrecision}\\
    \midrule
    $m_h$
    & --
    & $125.3\unit{GeV}$ ($2.40\unit{ms}$)
    & $124.1\unit{GeV}$ ($9.57\unit{ms}$) \\
    $m_{\tilde{g}}$
    & $2218\unit{GeV}$ ($0.12\unit{ms}$)
    & $2218\unit{GeV}$ ($0.12\unit{ms}$)
    & $2231\unit{GeV}$ ($0.40\unit{ms}$) \\
    $m_{\tilde{\chi}_1^0}$
    & $429\unit{GeV}$ ($4.02\unit{ms}$)
    & $429\unit{GeV}$ ($16.4\unit{ms}$)
    & $429\unit{GeV}$ ($48.3\unit{ms}$) \\
    $m_{\tilde{\chi}_4^0}$
    & $1943\unit{GeV}$ ($4.02\unit{ms}$)
    & $1944\unit{GeV}$ ($16.4\unit{ms}$)
    & $1944\unit{GeV}$ ($48.3\unit{ms}$) \\
    $m_{\tilde{u}_1}$
    & $1055\unit{GeV}$ ($2.50\unit{ms}$)
    & $1081\unit{GeV}$ ($15.1\unit{ms}$)
    & $1085\unit{GeV}$ ($59.1\unit{ms}$) \\
    $m_{\tilde{d}_1}$
    & $1801\unit{GeV}$ ($2.54\unit{ms}$)
    & $1778\unit{GeV}$ ($15.3\unit{ms}$)
    & $1783\unit{GeV}$ ($59.7\unit{ms}$) \\
    $m_{\tilde{e}_1}$
    & $1019\unit{GeV}$ ($1.90\unit{ms}$)
    & $1018\unit{GeV}$ ($11.3\unit{ms}$)
    & $1018\unit{GeV}$ ($22.6\unit{ms}$) \\
    \bottomrule
  \end{tabular}
  \caption{Comparison of pole masses with different calculation
    methods for the CMSSM with $m_0=M_{1/2}=1\unit{TeV}$, $A_0=-3.3\unit{TeV}$,
    $\tan\beta=10$, $\sign\mu=+1$.  In brackets the time used to calculate
    the pole masses of the whole multiplet is given.}
  \label{tab:pole-mass-precision-comparison}
\end{table}

For the Higgs states two-loop corrections to the self-energies can
optionally be added by setting \code{UseHiggs2LoopMSSM = True} in the
MSSM or \code{UseHiggs2LoopNMSSM = True} in the NMSSM in the model
file.  The former provides routines that call the two-loop Higgs
FORTRAN routines supplied by P.~Slavich from
\cite{Degrassi:2001yf,Brignole:2001jy,Dedes:2002dy,Brignole:2002bz,Dedes:2003km}
for calculating corrections of $\oatas$, $\oabas$, $\oatq$, $\oabq$,
$\oatauq$ and $\oatab$.  The latter adds corrections calculated in the
NMSSM at $\oatas$, $\oabas$ from \cite{Degrassi:2009yq} and partial
corrections for the order $\oatq$, $\oabq$, $\oatauq$ and $\oatab$
from the MSSM.  These corrections can then be included in the
calculation of the Higgs pole masses when running the spectrum
generator by setting the appropriate SLHA flags.

When two-loop corrections have been enabled in the SLHA file by
setting entry $4$ of the \code{FlexibleSUSY} block to $2$ the user may
also select individual corrections.  The \code{FlexibleSUSY} block
entries $9$, $10$, $11$ and $12$, correspond to two-loop corrections
of the order $\oatas$, $\oabas$, $\oatplusabsq$ and $\oatauq$
respectively and will be disabled when the corresponding entry is set
to zero.  In this way, for example, in the NMSSM the user may decide
not to use the partial corrections at order $\oatplusabsq$ and
$\oatauq$, all of which have only been computed in the MSSM.

Since the Higgs mass is a very important measurement and the two-loop
corrections can be larger than the current experimental error
\cite{Degrassi:2009yq} we recommend to set these switches to
\code{True} in any MSSM-\footnote{In \fs a model is MSSM-like if (i)
  its superpotential is approximately given by the MSSM one, (ii) it
  implements $R$-parity conservation or something equivalent, and
  (iii) contains two $CP$-even and $CP$-odd Higgs bosons, where one $CP$-odd
  Higgs boson may be a Goldstone boson, all with an MSSM-like coupling
  to $t$, $b$ and $\tau$.} and NMSSM-like\footnote{In \fs a model is
  NMSSM-like if (i) its superpotential is approximately given by the
  NMSSM one, (ii) it implements $R$-parity conservation or something
  equivalent, and (iii) contains three $CP$-even and $CP$-odd Higgs
  bosons, where one or two $CP$-odd Higgs bosons may be Goldstone
  bosons, all with an NMSSM-like coupling to $t$, $b$ and $\tau$\@.
  Examples for NMSSM-like models are the USSM and the \ESSM.} model.
However in such models the user should still consider whether these
corrections are really the leading corrections in the model or 
there are other potentially large two-loop corrections which are
missing.  For models with a more extended Higgs sector we recommend
that the leading log two-loop corrections are estimated by
generalizing those of the MSSM or NMSSM.

\section{Flexible Applications}
\label{sec:Flexible}

By definition, research is an endeavor to find something new.
Therefore, it can often be the case that
a spectrum generator right out of the box is not enough.
\fs attempts to offer a clean interface through which
one can exploit its facilities
while undergoing a minimal amount of frustration,
when one programs for a wide variety of studies.
We provide two basic levels for the user to create a custom spectrum
generator: (i) The \mathematica level, where one writes or
adapts a model file and (ii) the C++ level, where the generated
classes can be extended, recombined or replaced by self-made modules.
In what follows, adaptions on these two levels shall be demonstrated
by presenting a few use cases at differing degrees of complexity.

To avoid confusion,
it should be mentioned that
the code snippets presented below are not
verbatim listings of the files included in the package.
They have been tailored retaining the semantics
for conciseness.

\subsection{Adapting model files}
\label{sec:adapting-model-files}

There are simple but interesting goals
that one can achieve only by working on
\mathematica files.
The outcome thus obtained from \fs
might already include a fully-fledged program
that is useful in physics analysis.
In a more advanced project,
one might utilize the produced libraries as building blocks
that constitute the target application.
For a general account of the \fs model files,
refer to \secref{sec:modfile}.

\subsubsection{Changing boundary conditions}
\label{sec:changing boundary conditions}

As already emphasized in \secref{sec:Program}, the modular design of
\fs makes it straightforward 
to replace a boundary condition object.
The question then becomes how
one could obtain an alternative boundary condition class,
apart from writing one by hand.
The meta code feature of \fs offers great assistance
in this respect.
An example shall be presented to illustrate how this works.

In the literature,
there is a popular alternative to the CMSSM boundary condition
under which the Higgs soft masses are allowed to be different from
the universal mass of the other scalars \cite{NUHM}.
One might implement
this non-universal Higgs-mass MSSM (NUHMSSM) scenario
simply by modifying the model description given to \fs.
A section of the \code{FlexibleSUSY.m.in} file is listed below:
\begin{numlstlisting}
EXTPAR = {{1, mHd2In}, {2, mHu2In}};

HighScaleInput={
  {mHd2, mHd2In}, {mHu2, mHu2In},
  {T[Ye], Azero*Ye}, {T[Yd], Azero*Yd}, {T[Yu], Azero*Yu},
  {mq2, UNITMATRIX[3] m0^2}, {ml2, UNITMATRIX[3] m0^2}, {md2, UNITMATRIX[3] m0^2},
  {mu2, UNITMATRIX[3] m0^2}, {me2, UNITMATRIX[3] m0^2},
  {MassB, m12}, {MassWB, m12}, {MassG, m12}
};
\end{numlstlisting}
Since \code{mHd2} and \code{mHu2} are to be fixed at constants
different from \code{m0^2},
two additional input parameters,
\code{mHd2In} and \code{mHu2In}, holding those constants,
are introduced in the list \code{EXTPAR}.
These input parameters are then declared to be the high-scale values of
\code{mHd2} and \code{mHu2} in line 4.
The rest of the boundary conditions is the same as in the CMSSM\@.
In the SLHA input file,
the parameter indices \code{1} and \code{2} of
\code{mHd2In} and \code{mHu2In}, declared in \code{EXTPAR} above,
must appear
as the first field in each line in the \code{EXTPAR} block:
\begin{numlstlisting}
Block EXTPAR
    1   10000                # mHd2In
    2   -2500                # mHu2In
\end{numlstlisting}
Note that the two additional input parameters are chosen to have
mass dimension 2, unlike \code{m0}.
This makes it easy to
try both signs of the high-scale value of either soft Higgs mass squared,
as exemplified in line 3.
If one were not interested in a negative boundary value of \code{mHu2}
for instance,
then a dimension-1 parameter 
might instead be introduced whose square is equated with \code{mHu2}.

The full implementation is available
in \code{model_files/NUHMSSM/}.
To try it out, do the following:
\begin{numlstlisting}[language=bash]
$ ./createmodel --name=NUHMSSM --sarah-model=MSSM
$ ./configure --with-models=NUHMSSM
$ make
$ models/NUHMSSM/run_NUHMSSM.x --slha-input-file=models/NUHMSSM/LesHouches.in.NUHMSSM
\end{numlstlisting}
Notice the \code{--sarah-model=MSSM} flag in line 1.
It tells the \code{createmodel} script to reuse
the MSSM specification in \sarah
to generate the C++ program.
Another remark is in order regarding the naming convention
of specimen SLHA input files.  The \code{createmodel} script assumes
that their names are in the form \code{LesHouches.in*} (case-insensitive).
If the script finds such files in \code{model_files/<model>/},
it installs them into the model directory.  The argument to
\code{--slha-input-file=} in line 4 has been thus created.

\subsubsection{Extending existing models}
\label{sec:extending existing models}

The preceding example was a modest alteration of a physics scenario
in that an existing model has been reused.
A more non-trivial modification might involve
an extension of the particle content as well as the interactions.
One of the simplest classes of models beyond the MSSM is
those with additional gauge-singlet fields.
In what follows, a supersymmetric type-I see-saw scenario
\cite{see-saw} 
shall be considered.
For this, two extensions of the MSSM are introduced:
MSSMRHN with three extra neutral (heavy) chiral superfields,
and MSSMD5O with the dimension-5 neutrino mass operator
added to the superpotential.
Both models are included in the package.

The name MSSMRHN of the first model stands for
the MSSM plus right-handed neutrinos.
One needs to prepare an input file to \sarah which might be placed in
\code{<FlexibleSUSY-root>/sarah/MSSMRHN/} or
\code{<SARAH-root>/Models/MSSMRHN/}.
The input file \code{MSSMRHN.m} contains
the declaration of the three-generation singlets \code{v}:
\begin{numlstlisting}[name=MSSMRHN.m]
SuperFields[[8]] = {v, 3, conj[vR], 0, 1, 1, RpM};
\end{numlstlisting}
as well as the neutrino Yukawa couplings and the Majorana mass terms
of the singlets:
\begin{numlstlisting}[name=MSSMRHN.m]
SuperPotential = Yu u.q.Hu - Yd d.q.Hd - Ye e.l.Hd + \[Mu] Hu.Hd +
                 Yv v.l.Hu + Mv/2 v.v;
\end{numlstlisting}
Further declarations inform \sarah
of how to form Dirac spinors out of the new Weyl spinors
and how the scalars and the fermions mix to comprise
the mass eigenstates:
\begin{numlstlisting}[name=MSSMRHN.m]
DEFINITION[GaugeES][DiracSpinors] = {
  Fu1 -> {FuL, 0}, Fu2 -> {0, FuR},
  Fv1 -> {FvL, 0}, Fv2 -> {0, FvR},
  ...
};

DEFINITION[EWSB][MatterSector] = {
  {{SuL, SuR}, {Su, ZU}},
  {{SvL, SvR}, {Sv, ZV}},
  ...
  {{fB, fW0, FHd0, FHu0}, {L0, ZN}},
  {{FvL, conj[FvR]}, {FV, UV}},
  {{{fWm, FHdm}, {fWp, FHup}}, {{Lm, UM}, {Lp, UP}}},
  {{{FuL}, {conj[FuR]}}, {{FUL, ZUL}, {FUR, ZUR}}}
};

DEFINITION[EWSB][DiracSpinors] = {
  Fu  -> {FUL, conj[FUR]},
  Fv  -> {FV , conj[FV] },
  Chi -> {L0 , conj[L0] },
  Cha -> {Lm , conj[Lp] },
  ...
};
\end{numlstlisting}
With respect to the MSSM file, the newly added lines are
6, 12, 15, and 22.
Notice that the (left- and right-handed) neutrino mixing
in line 15 resembles the neutralino mixing in line 14.
Due to the Majorana mass term in the superpotential,
the six neutrino mass eigenstates are described in terms of
Majorana spinors like the neutralinos.

One should then add descriptions of the new states in the file
\code{particles.m}:
\begin{numlstlisting}
ParticleDefinitions[GaugeES] = {
  {Fv1, { Description -> "Dirac Left Neutrino" }},
  {Fv2, { Description -> "Dirac Right Neutrino" }},
  {SvR, { Description -> "Right Sneutrino", LaTeX ->"\\tilde{\\nu}_R"}},
  ...
};

ParticleDefinitions[EWSB] = {
  {Sv, { Description -> "Sneutrinos",
         PDG -> {1000012, 1000014, 1000016, 2000012, 2000014, 2000016}}},
  {Fv, { Description -> "Neutrinos",
         PDG -> {12, 14, 16, 9900012, 9900014, 9900016}}},
  ...
};

WeylFermionAndIndermediate = {
  {v,   { Description -> "Right Neutrino Superfield" }},
  {FV,  { Description -> "Neutrino-Masseigenstate"}},
  {FvL, { Description -> "Left Neutrino"}},
  {FvR, { Description -> "Right Neutrino"}},
  ...
};
\end{numlstlisting}
In line 12, one finds PDG codes beginning with \code{99}.
Such numbers are available for a program author's private use
\cite{Beringer:1900zz}.
The new parameters in the superpotential
and the soft supersymmetry breaking sector
are to be described in \code{parameters.m}:
\begin{numlstlisting}
ParameterDefinitions = {
  {UV,    { Description -> "Neutrino-Mixing-Matrix"}},
  {Yv,    { Description -> "Neutrino-Yukawa-Coupling" }},
  {T[Yv], { Description -> "Trilinear-Neutrino-Coupling"}},
  {Mv,    { LaTeX -> "M_v", OutputName -> Mv, LesHouches -> Mv}},
  {B[Mv], { LaTeX -> "B_v", OutputName -> BMv, LesHouches -> BMv}},
  {mv2,   { Description -> "Softbreaking right Sneutrino Mass"}},
  ...
};
\end{numlstlisting}
For further details on how to write model files for \sarah, we refer
to its manual \cite{Staub:2008uz,Staub:2013tta}.

Finally, it remains to put \code{FlexibleSUSY.m.in}
in \code{model_files/MSSMRHN/}.
The high-scale boundary conditions therein might read:
\begin{numlstlisting}
HighScaleInput = {
  {mv2, UNITMATRIX[3] m0^2},
  {T[Yv], Azero*Yv},
  {B[Mv], LHInput[B[Mv]]},
  ...
};
\end{numlstlisting}

The second model is called MSSMD5O, standing for
the MSSM including the dimension-5 operator.
Obviously, one can compose it by adding
the additional term to the superpotential in \code{MSSMD5O.m}:
\begin{numlstlisting}
SuperPotential = Yu u.q.Hu - Yd d.q.Hd - Ye e.l.Hd + \[Mu] Hu.Hd \
               + WOp/2 l.Hu.l.Hu;
\end{numlstlisting}
where line 2 contains the dimension-5 operator
multiplied by its coefficient matrix \code{WOp}.
The declarations of the neutrino Dirac spinors and mixing are
very similar to those in the MSSMRHN,
except that \code{FvR} is absent.

One can specify the low-scale constraints on \code{WOp}
in \code{model_files/MSSMRHN/FlexibleSUSY.m.in}:
\begin{numlstlisting}
EXTPAR = {
  {1, mv1}, {2, mv2}, {3, mv3},
  {4, ThetaV12}, {5, ThetaV13}, {6, ThetaV23},
  {7, YvDiag1}, {8, YvDiag2}, {9, YvDiag3}
};

UPMNS = Module[{
    s12 = Sin @ ThetaV12, c12 = Cos @ ThetaV12,
    s13 = Sin @ ThetaV13, c13 = Cos @ ThetaV13,
    s23 = Sin @ ThetaV23, c23 = Cos @ ThetaV23
  },
  {{  c12 c13              ,  s12 c13              , s13     },
   { -s12 c23 - c12 s23 s13,  c12 c23 - s12 s23 s13, s23 c13 },
   {  s12 s23 - c12 c23 s13, -c12 s23 - s12 c23 s13, c23 c13 }}
];

mv = conj[UPMNS].DiagonalMatrix[{mv1, mv2, mv3}].Transpose[conj @ UPMNS];

LowScaleInput = Join[
  { (* MSSM low-scale constraints *) },
  Flatten[Table[{WOp[i,j], mv[[i,j]] / (vu/Sqrt[2])^2}, {i,3}, {j,3}], 1]
];

InitialGuessAtLowScale = Join[
  { (* MSSM initial guesses at low scale *) },
  Flatten[Table[{WOp[i,j], mv[[i,j]] / (vu/Sqrt[2])^2}, {i,3}, {j,3}], 1]
];
\end{numlstlisting}
The \code{EXTPAR} list contains the input parameters
to be read from the corresponding SLHA block.
The low-energy neutrino mass eigenvalues and mixing angles are
declared in lines 2--3.
They are followed by the neutrino Yukawa eigenvalues,
which shall be used as part of the matching condition
described in \ref{sec:tower construction}.
The constraint on and the initial guess of \code{WOp}
in lines 21 and 26 should be self-explanatory.

With the above set of input files,
\fs can generate the C++ class libraries,
\code{libMSSMRHN} and \code{libMSSMD5O}\@.
These products shall be employed as the two effective theories
in the implementation of the see-saw mechanism.
To this end, one further needs to code at the C++ level,
as explained in the next subsection and \ref{sec:tower construction}.

\subsection{Adapting C++ code}
\label{sec:adapting-cpp-code}

There are problems which one cannot solve only by
editing \mathematica model files.
To unlock the full potential of \fs,
it is an advantage not to avoid programming at the C++ level.
For this, it should help to have working knowledge about the basic structure
of a spectrum generator, set out in \secref{sec:SpecGenStruct}.
In \ref{sec:examples of c++ code adaptation},
two examples are presented for demonstrating that
the clean class structure serves as firm guidance on the job.

The first project in \ref{sec:tower construction} is
to build a spectrum generator that can handle
a tower of multiple effective field theories.
The aim is to take a first step towards
a study of slepton-mediated lepton flavour violation
due to radiative corrections in the type-I supersymmetric see-saw model
\cite{Borzumati:1986qx}.
To this end, \code{MSSMRHN} is stacked on top of \code{MSSMD5O}.
The preparation of these two models has been covered in
\secref{sec:extending existing models}.

Since each model class has its own $\beta$-functions,
the spectrum generator contains two different sets of RGEs
that are connected by a matching object.
The program shall accept
the low-energy neutrino masses and mixing angles
which determine \code{WOp}, the coefficients
of the $L H_u L H_u$ operator.
These $6 \times 6$ coefficients evolve to
the right-handed neutrino mass scale at which
they are matched to the neutrino Yukawa couplings \code{Yv} and
the right-handed neutrino masses \code{Mv}.
Since there are more degrees of freedom in the pair of
\code{Yv} and \code{Mv} than in \code{WOp},
one needs supplementary conditions in addition to the see-saw relation.
In the presented matching code, it is assumed that all mixing in \code{WOp}
stems from the left-handed rotation of \code{Yv} whose eigenvalues are
fixed to those specified by the user.
The non-trivial flavour structure of the neutrino Yukawa couplings then
causes running slepton mass matrices to acquire flavour-violating elements.
The output from the spectrum generator includes
the slepton mass matrices as well as
the resulting mass eigenvalues and mixing.
One might pass this outcome on to another routine
to calculate rates of lepton flavour violating processes.

For brevity,
threshold corrections are ignored
in the specimen matching code connecting the two models
as well as in the low-scale boundary condition on \code{WOp}
from the neutrino oscillation data.
Therefore, the result maintains only the accuracy of one-loop RGEs,
even though two-loop $\beta$-functions are computed by each model class.
For a full accuracy of two-loop RGEs, one can incorporate the omitted
one-loop corrections into the constraint classes.
The way to implement them should be self-evident from the code structure.

The second project in \ref{sec:integrating-custom-built} shows
how one can employ new spectrum generator components,
which may be composed from scratch or through
a linkage to external routines.
In the procedure,
it would be noticed that there is an evident limit on the scope of
modules which one has to deal with.
For instance, it is clear from the outset
that one does not have to go through the code of the central
fixed-point iteration engine, \code{RGFlow}.
This manifests the power of the clear separation among objects
each with its well-defined distinct role.
This is just like the fact that one does not need to access the internals
of the \code{std::sort} function in the C++ Standard Library.
It might be entertaining to complete the analogy by
mapping the model objects in \code{RGFlow}
to the elements that \code{std::sort} sorts
and the boundary condition objects
to the comparator function.

\section{Tests and comparisons with other spectrum generators}
\label{sec:comparison}

\subsection{Numeric tests}

To check the correctness of \fs's generated spectrum generators
extensive unit testing against Softsusy's MSSM and NMSSM
implementations (both $Z_3$-invariant and $Z_3$-violating variants)
has been carried out.  These unit tests systematically compare all
tree-level mass matrices, EWSB equations, one- and two-loop
$\beta$-functions, one- and two-loop self-energies and one- and
two-loop tadpoles numerically for the CMSSM, the semi-constrained
$Z_3$-invariant NMSSM ($Z_3$-NMSSM)\footnote{With the semi-constrained
  $Z_3$-invariant NMSSM ($Z_3$-NMSSM) we denote a constrained variant
  of the NMSSM with universal gaugino masses $M_{1/2}$, universal
  trilinear couplings $A_0$ and universal MSSM-like soft-breaking
  squared scalar masses $m_0^2$ at the GUT scale.  The soft-breaking
  singlet mass $m_S^2$, the trilinear singlet superpotential coupling
  $\kappa$ and the singlet VEV $s$ are fixed by the EWSB conditions at
  the SUSY scale.  The $Z_3$-NMSSM has the $6$ free parameters
  $(m_0^2, M_{1/2}, A_0, \tan\beta, \sign\mu_\text{eff}, \lambda)$.}
and the constrained $Z_3$-violating NMSSM
($\Zv_3$-NMSSM)\footnote{With the constrained $Z_3$-violating NMSSM
  ($\Zv_3$-NMSSM) we denote a constrained variant of the
  $Z_3$-violating NMSSM with universal gaugino masses $M_{1/2}$,
  universal trilinear couplings $A_0$ and universal MSSM-like
  soft-breaking squared scalar masses $m_0^2$ at the GUT scale.  The
  $\mu$-parameter, its soft-breaking equivalent $B\mu$ and the
  soft-breaking singlet tadpole coupling $\xi_S$ are fixed by the EWSB
  conditions at the SUSY scale.  The $\Zv_3$-NMSSM has the $11$ free
  parameters $(m_0^2, M_{1/2}, A_0, \tan\beta, \sign\mu_\text{eff},
  \lambda, \kappa, s, \mu', m_S^{\prime 2}, \xi_F)$.}  parameter
points given in \tabref{tab:unit-test-parameter-points}.
\begin{table}[tbh]
  \centering
  \begin{tabularx}{\textwidth}{lX}
    \toprule
    SUSY Model & Tested parameter points\\
    \midrule CMSSM & $m_0 = 125\unit{GeV}$, $M_{1/2} = 500\unit{GeV}$,
    $\tan\beta = 10$, $A_0 = 0$, $\sign\mu = \pm 1$
    \\
    $Z_3$-NMSSM & $m_0 = \{250,300\}\unit{GeV}$, $M_{1/2} =
    200\unit{GeV}$, $\tan\beta = 10$, $A_0 = -500\unit{GeV}$,
    $\sign\mu_\text{eff} = +1$, $\lambda = 0.1$
    \\
    $\Zv_3$-NMSSM & $m_0 = 540\unit{GeV}$, $M_{1/2} = 200\unit{GeV}$,
    $\tan\beta = 10$, $A_0 = -350\unit{GeV}$, $\sign\mu_\text{eff} =
    \pm 1$, $\lambda = \kappa = 0.1$, $s = 1\unit{TeV}$, $\mu' =
    290\unit{GeV}$, $m_S^{\prime 2} = 400\unit{GeV}$, $\xi_F =
    300\unit{GeV}$
    \\
    \bottomrule
  \end{tabularx}
  \caption{CMSSM, semi-constrained
    $Z_3$-invariant NMSSM ($Z_3$-NMSSM) and constrained
    $Z_3$-violating NMSSM ($\Zv_3$-NMSSM) parameter points
    used for the unit tests against Softsusy.
    We follow the notation of the NMSSM model parameters used in
    \cite{Allanach:2013kza,Ellwanger:2009dp}.}
  \label{tab:unit-test-parameter-points}
\end{table}
All tested expressions were found to agree within double machine
precision.\footnote{Due to the systematic and detailed tests several
  bugs in Softsusy, SARAH and \fs could be identified and corrected.}
Furthermore, the output of the iterative procedures which solve the
one- and two-loop corrected tadpole equations
\eqref{eq:one-loop-ewsb-eq} to find the minimum of the effective Higgs
potential were compared numerically for these parameter points and
found to agree within machine precision as well.  Finally, the overall
pole mass spectrum and mixing after the full fixed-point iteration has
finished has been compared, and was found to agree at the sub-permille
level.  The origin of the sub-permille level difference between \fs
and Softsusy is the different determination of the weak mixing angle
$\theta_{W,\text{susy}}^{\text{\DRbar}}(M_Z)$ in the SUSY model in the
\DRbar\ scheme: \fs calculates
$\theta_{W,\text{susy}}^{\text{\DRbar}}(M_Z)$ from $M_W$ and $M_Z$, as
described in \secref{sec:calculation-of-gauge-couplings}, while
Softsusy determines $\theta_{W,\text{susy}}^{\text{\DRbar}}(M_Z)$ from
the muon decay constant $G_\mu$.  The approach used in Softsusy
results in more precise \DRbar\ gauge couplings at the $M_Z$ scale,
because the muon decay constant is known with a higher accuracy than
the $W$-boson mass.  Furthermore, in Softsusy $3.5.0$ some three-loop
$\beta$-functions and two-loop threshold corrections can be enabled in
the MSSM to increase the accuracy of the RG running and the
determination of the \DRbar\ gauge and Yukawa couplings at $M_Z$
\cite{Allanach:2014nba}.  These corrections are not implemented in \fs
so far.  The complete set of unit tests is shipped with \fs and can be
found in the \code{test/} directory.  The tests can be run with the
command \code{make execute-tests}.  All unit tests are carried out
nightly in order to continuously check the correctness of the meta
code and the generated spectrum generators for the shipped models.
The nightly test results can be found at
\url{https://www.desy.de/~alvoigt/FlexibleSUSY/test.xhtml}

The \fs generated NUHM \ESSM\ spectrum generator has also been
compared against a handwritten one for a constrained version of the
\ESSM \cite{Athron:2009ue, Athron:2009bs, Athron:2012pw}.  The
$\beta$-functions were systematically compared in unit tests and were
found to match within numerical precision.  The handwritten code
does not include full one-loop self-energies or tadpoles, so tests on
these were not carried out.  Although the generators assume different
constraints and solve the boundary value problem with completely
different algorithms they could be compared by using the output of
the CE$_6$SSM generator as an input to \fs and the spectra were found
to be in reasonable agreement  given the different levels of precision with deviations in the mass spectra $\lesssim 10\%$.

In addition \fs has already undergone some user testing. This includes
analytic tests of the $R$-symmetric low-energy model (MRSSM) and
alternative $E_6$-inspired SUSY scenarios.  The users who have helped us
with this are thanked in the acknowledgements.

We also compared the run-time of \fs against SPheno, Softsusy and the
SARAH generated MSSM spectrum generator SPhenoMSSM.  The results of
the comparison can be found in \secref{sec:run-time comparison}.

\subsection{Run-time comparison}
\label{sec:run-time comparison}

One of \fs's design goals is a short run-time.  In this section we
demonstrate that this goal was achieved by comparing the run-time of
two different sets of CMSSM spectrum generators:
\begin{itemize}
\item \emph{Without sfermion flavour violation:} Disallowing sfermion
  flavour violation simplifies the calculation of the pole masses,
  because flavour-off-diagonal sfermion self-energy matrix elements
  do not need to be calculated.  Here we compare \fs's non-flavour
  violating CMSSM spectrum generator FlexibleSUSY-NoFV (version 1.0.0)
  against SPheno (version 3.2.4) and Softsusy (version 3.4.0).
\item \emph{With sfermion flavour violation:} Allowing for sfermion
  flavour violation in general increases the run-time of spectrum
  generators, because the full $6\times 6$ sfermion self-energy
  matrices have to be calculated.  Here we compare FlexibleSUSY-FV
  (version 1.0.0) and SPhenoMSSM (generated with SARAH 4.1.0 and
  linked against SPheno 3.2.4).  Both spectrum generators are based on
  SARAH's MSSM model file, which allows for sfermion flavour
  violation.
\end{itemize}
For the run-time comparison the CKM matrix is set to unity, all
$CP$-violating phases are set to zero and $R$-parity violation is
disabled.  \fs and Softsusy are compiled with g++ 4.8.0 and Intel
ifort 13.1.3 20130607.  SPheno and SPhenoMSSM are compiled with Intel
ifort 13.1.3 20130607.\footnote{Intel's ifort compiler decreases the
  run-time of SPheno and SPhenoMSSM by approximately a factor $1.5$,
  compared to gfortran.}  We are generating $2\cdot 10^{4}$ random
CMSSM parameter points with $m_0\in [50,1000]\unit{GeV}$, $m_{1/2}\in
[50,1000]\unit{GeV}$, $\tan\beta\in [1,100]$, $\sign\mu\in \{-1,+1\}$
and $A_0\in [-1000,$ $1000]\unit{GeV}$.  For each point an SLHA input
file is created by appending the values of $m_0$, $m_{1/2}$,
$\tan\beta$, $\sign\mu$, $A_0$ in form of a \code{MINPAR} block to the
SLHA template file given in \ref{sec:speed-test-slha-template-file}.
The resulting SLHA input file is passed to each spectrum generator and
the (wall-clock) time is measured until the program has finished.  The
average run-times for three different CPU types can be found in
\tabref{tab:run-time-comparison}.  The first column shows the run-time
on an Intel Core2 Duo (P8600, $2.40\unit{GHz}$) where only one core was
enabled.  The second column shows the run-time on the same processor
where both cores were enabled.  In the third column a machine with two
Intel Xeon CPUs (L5640, $2.27\unit{GHz}$, $6$ cores) was used.
\begin{table}[tbh]
  \centering
  \begin{tabular}{llll}
    \toprule
                            & Intel Core2 Duo    & Intel Core2 Duo   & $2$ $\times$ Intel Xeon\\
                            & (P8600, $1$ core)  & (P8600, $2$ cores)& (L5640, $6$ cores)\\
    \midrule
    FlexibleSUSY-NoFV 1.0.0 & $0.086\unit{s}$    & $0.079\unit{s}$   & $0.060\unit{s}$\\
    SPheno 3.2.4            & $0.119\unit{s}$    & $0.114\unit{s}$   & $0.101\unit{s}$\\
    Softsusy 3.4.0          & $0.175\unit{s}$    & $0.171\unit{s}$   & $0.147\unit{s}$\\
    \midrule
    FlexibleSUSY-FV 1.0.0   & $0.150\unit{s}$    & $0.113\unit{s}$   & $0.074\unit{s}$\\
    SPhenoMSSM 4.1.0        & $0.415\unit{s}$    & $0.401\unit{s}$   & $0.370\unit{s}$\\
    \bottomrule
  \end{tabular}
  \caption{Average run-time of CMSSM spectrum generators
    for random parameter points.  The first three rows show three
    spectrum generators with disabled sfermion flavour violation.  Rows
    $4$--$5$ show two spectrum generators with enabled sfermion flavour violation, both based
    on SARAH's MSSM model file.}
  \label{tab:run-time-comparison}
\end{table}

Among both the non-flavour violating spectrum generators (first three
rows) as well as the flavour violating ones (4th and 5th row) we find
that \fs is significantly fastest.  Compared to SPheno,
FlexibleSUSY-NoFV is faster by a factor $1.4$--$1.7$, and compared to
Softsusy around a factor $2$--$2.5$.  Between the flavour violating
spectrum generators FlexibleSUSY-FV is faster than SPhenoMSSM by a
factor $2.8$--$5$.  Reason for the long run-time of SPhenoMSSM is the
long calculation duration of the two-loop $\beta$-functions.  Here \fs
benefits a lot from Eigen's well-optimizable matrix expressions.  We
also find that increasing the number of CPU cores reduces the run-time
of \fs.  The reason is that \fs calculates each pole mass in a
separate thread, and therefore benefits from multi-core CPUs.

\section{Conclusions}

We have presented \fs, a Mathematica and C++ package, which generates fast and
modular spectrum generators for any user specified SUSY model.  \fs is
distributed with a large number of predefined models for the CMSSM,
NMSSM, USSM, \ESSM, MRSSM etc., which can be generated immediately
without any editing.  In particular the CMSSM and NMSSM spectrum
generators constitute a fast and reliable alternative to the existing
publicly available spectrum generators, Softsusy, SPheno and
NMSPEC. 

We have described how the generated source code can be influenced at
two different levels: The \mathematica level where the user provides a
model file, and the C++ level where the generated objects can be
easily exchanged, extended, modified and reused.  This provides great
flexibility for creating custom spectrum generators for both the most
common and most extraordinary models.  We have demonstrated these
features in detailed examples for the NUHMSSM, right handed neutrinos
and on adding three-loop RGEs and two-loop matching for the strong
gauge coupling.  

The generated code has been extensively tested against Softsusy, and
additional tests have been carried out for non-minimal models, the
\ESSM and MRSSM.  Speed tests have also been performed against
Softsusy, SPheno and SPheno-like MSSM code generated by SARAH,
demonstrating that \fs runs faster than all three.

As a result \fs enables fast exploitation of new SUSY models with high
precision and reliability.

\section*{Acknowledgments}

A.V.\ would like to thank Florian Staub for countless explanations of
SARAH's internals, discussions, and exceptionally fast bug
fixing. P.A.~would like to thank Roman Nevzorov for useful discussions
about challenges in non-minimal SUSY models and the GAMBIT
collaboration for helpful suggestions, particularly regarding
lowering our compiler prerequisites. The authors would also like to
thank Lewis Tunstall for helping with early tests against
Next-to-Minimal Softsusy; Sophie Underwood for discovering problems
when introducing couplings with a single family index; Gregor Hellwig
for writing the first version of the E6SSM model file; Philip Diessner
for further adaptions to the E6SSM model file and for supplying it to
us and for for testing and identifying several bugs in \fs and in
SARAH, in work on the MRSSM; Ulrik Günther for compilation tests on
Mac OS X and Dylan Harries for spotting a bug in the configuration
script and creating an EWSB fixed-point iteration implementation for
the CNE$_6$SSM, which served as a prototype for implementing a
generalized EWSB fixed-point iteration.
J.P. acknowledges support from the MEC and FEDER (EC) Grants
FPA2011--23596 and the Generalitat Valenciana under grant PROMETEOII/2013/017.
This work has been supported by the German Research Foundation DFG
through Grant No.~STO876/2-1.

\appendix

\section{Examples of C++ code adaptation}
\label{sec:examples of c++ code adaptation}

In what follows, technical details of \fs programming at the C++ level are
set out which supplement the outline given
in \secref{sec:adapting-cpp-code}.

\subsection{Stacking models in a tower of effective theories}
\label{sec:tower construction}

\begin{figure}
  \centering
\begin{tikzpicture}
  \draw (0,-.5) coordinate(bottom)
        (0,-.4)
     -- (0,0)  coordinate(MZ) node[left] {$M_Z$}
     -- (0,.7) coordinate(MS) node[left] {$M_S$}
     -- (0,2)  coordinate(Mv) node[left] {$M_\nu$}
     -- (0,3)  coordinate(MX) node[left] {$M_X$}
     -- (0,3.4)
        (0,3.5) coordinate(top);

  \path (MZ) ++(.3,0) coordinate(MZR) +(3,0) coordinate(MZRR)
        (MS) ++(.3,0) coordinate(MSR) +(3,0) coordinate(MSRR)
        (Mv) ++(.3,0) coordinate(MvR) +(3,0) coordinate(MvRR)
        (MX) ++(.3,0) coordinate(MXR) +(3,0) coordinate(MXRR);

  \draw (MZ) +(-.1,0) -- +(+.1,0); \draw[<-] (MZ) ++(-.8,0) -- ++(-.5,0)
     node[left,draw] {\code{MSSMD5O_low_scale_constraint}};
  \draw (MS) +(-.1,0) -- +(+.1,0); \draw[<-] (MS) ++(-.8,0) -- ++(-.5,0)
     node[left,draw] {\code{MSSMD5O_susy_scale_constraint}};
  \draw (Mv) +(-.1,0) -- +(+.1,0); \draw[<-] (Mv) ++(-.8,0) -- ++(-.5,0)
     node[left,draw] {\code{MSSMD5O_MSSMRHN_matching}};
  \draw (MX) +(-.1,0) -- +(+.1,0); \draw[<-] (MX) ++(-.8,0) -- ++(-.5,0)
     node[left,draw] {\code{MSSMRHN_high_scale_constraint}};

  \draw (MvRR) +(0,-.03) rectangle (MZR) node[pos=.5] {\code{MSSMD5O}};
  \draw (MvRR) +(0, .03) rectangle (MXR) node[pos=.5] {\code{MSSMRHN}};

  \path (bottom) +(3.8,0) coordinate(SE) (top) +(-6.8,0) coordinate(NW);
  \draw  (NW) +(0,.5) rectangle (SE)
         (NW) -- +(10.6,0);
  \path (top) node[above] {\code{RGFlow}};
\end{tikzpicture}
  \caption{Schematic class structure in the C++ code for
    the tower scenario.}
  \label{fig:class structure for tower}
\end{figure}
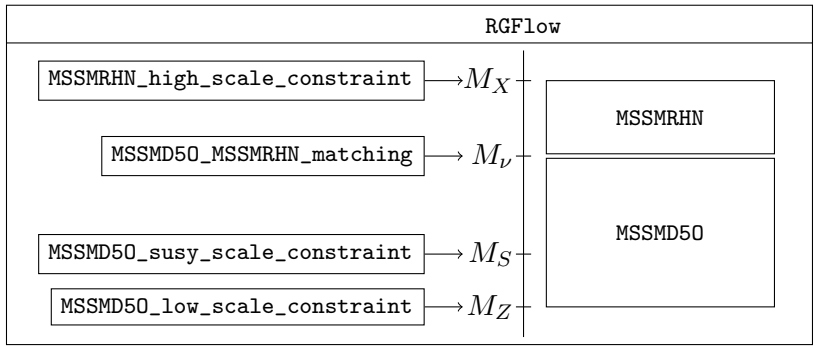

Consider a physics scenario which is best described
by a tower of effective theories.
Within the framework of \fs, the C++ class structure is
a faithful reflection of
this physicist's view on the given problem.
Here we illustrate this point
using a well-known configuration in which
the higher-energy theory is the MSSMRHN which
gives rise to the MSSMD5O as the lower-energy effective theory.
The relevant classes are sketched
in \figref{fig:class structure for tower}.
The \code{MSSMRHN} object is in effect from the $M_X$ scale down to
the $M_\nu$ scale at which the right-handed neutrinos are decoupled.
Below this scale, the \code{MSSMD5O} object takes over.
On the left of the vertical axis,
the boundary condition objects acting on either model are displayed,
together with the matching object connecting the two theories.
Note that each of the boundary condition and matching objects
maintains and updates its own scale over iterations.
An arrow in the figure depicts
the association of a constraint with its scale.
All these components are plugged into the \code{RGFlow} object
which then solves the problem.

The matching class as well as gluing codes have to be written by hand
to build such a program.\footnote{It is planned that a future release
  of \fs will be capable of creating this code automatically.}  All
remaining components of a multi-model spectrum generator can be authored
by making a straightforward extension
to each corresponding single-model counterpart
for one of the models forming the tower.

As the target spectrum generator depends on two models,
one should first build these prerequisites by:
\begin{numlstlisting}
$ ./createmodel --name=MSSMD5O
$ ./createmodel --name=MSSMRHN
$ ./configure --with-models=MSSMD5O,MSSMRHN
$ make
\end{numlstlisting}
As a by-product, line 3
also creates a \code{Makefile} in \code{examples/tower/}.
One can best see the overall code structure of the application in this file:
\begin{numlstlisting}
CPPFLAGS  := -I. $(INCCONFIG) $(INCFLEXI) $(INCLEGACY) $(INCSLHAEA) \
             $(INCMSSMD5O) $(INCMSSMRHN)

TOWER_SRC := run_tower.cpp \
	     MSSMD5O_MSSMRHN_two_scale_matching.cpp \
	     MSSMD5O_MSSMRHN_two_scale_initial_guesser.cpp

TOWER_OBJ := $(patsubst %.cpp, %.o, $(filter %.cpp, $(TOWER_SRC)))

run_tower.x: $(TOWER_OBJ) $(LIBMSSMD5O) $(LIBMSSMRHN) $(LIBFLEXI) $(LIBLEGACY)
  $(CXX) -o $@ $^ $(LOOPFUNCLIBS) $(GSLLIBS) $(BOOSTTHREADLIBS) $(THREADLIBS) $(LAPACKLIBS) $(BLASLIBS) $(FLIBS)
\end{numlstlisting}
The include directives in line 2 tell the compiler
where to find the headers for either \code{MSSMD5O} or \code{MSSMRHN}.
The \code{.cpp} files in lines 4--6 and the \code{.hpp} files
that they include are to be written by hand.
Obviously, the executable \code{run_tower.x}, in line 10, depends on both
\code{$(LIBMSSMD5O)} and \code{$(LIBMSSMRHN)}
that implement the auto-generated
components in \figref{fig:class structure for tower}.

To prepare the main source file \code{run_tower.cpp},
one can extend 
\code{run_MSSMD5O.cpp} or \code{run_MSSMRHN.cpp}
produced in either model directory.
The shipped example reads:
\begin{numlstlisting}
#include "MSSMD5O_MSSMRHN_spectrum_generator.hpp"

int main(int argc, char* argv[])
{
  // define objects;
  QedQcd oneset;
  MSSMD5O_input_parameters input_1;
  MSSMRHN_input_parameters input_2;
  // fill in input_1 and input_2;
  oneset.toMz(); // run SM fermion masses to MZ
  typedef Two_scale algorithm_type;
  MSSMD5O_MSSMRHN_spectrum_generator<algorithm_type> spectrum_generator;
  // set up spectrum_generator;
  spectrum_generator.run(oneset, input_1, input_2);
  // extract outcome from models;
}
\end{numlstlisting}
where a line in the form \code{// ...;} shall be understood
to be a pseudo-code.
Given two models,
one declares two sets of input parameters,
\code{input_1} and \code{input_2}, in lines 7--8.

The crucial point is the definition of the
\code{MSSMD5O_MSSMRHN_spectrum_generator} object in line 12, which
creates and drives the \code{RGFlow} object in \figref{fig:class
  structure for tower}.  This task is started by calling the
\code{run()} member function in line 14.  It is defined in
\code{MSSMD5O_MSSMRHN_spectrum_generator.hpp} and reads:
\begin{numlstlisting}[name=SGrun,language=C++]
template<class T> void MSSMD5O_MSSMRHN_spectrum_generator<T>::run
(const QedQcd& oneset,
 const MSSMD5O_input_parameters& input_1, const MSSMRHN_input_parameters& input_2)
{
  high_scale_constraint_2.clear(); // of type MSSMRHN_high_scale_constraint<T>
  susy_scale_constraint_1.clear(); // of type MSSMD5O_susy_scale_constraint<T>
  low_scale_constraint_1 .clear(); // of type MSSMD5O_low_scale_constraint<T>
  matching.reset();                // of type MSSMD5O_MSSMRHN_matching<T>
  high_scale_constraint_2.set_input_parameters(input_2);
  susy_scale_constraint_1.set_input_parameters(input_1);
  low_scale_constraint_1 .set_input_parameters(input_1);
  matching.set_lower_input_parameters(input_1);
  high_scale_constraint_2.initialize();
  susy_scale_constraint_1.initialize();
  low_scale_constraint_1 .initialize();
  if (!is_zero(input_scale_2)) high_scale_constraint_2.set_scale(input_scale_2);
\end{numlstlisting}
This piece of code is nearly a verbatim copy of
the corresponding part of \code{MSSMD5O_spectrum_generator.hpp}.
The only differences are that the type of
\code{high_scale_constraint_2} is
\code{MSSMRHN_high_scale_constraint<T>} and that
the \code{matching} object has been added.
Recall that the template parameter \code{T} has been bound to
\code{Two_scale} in the \code{main} function.
One then constructs a list of
the constraints on \code{MSSMD5O}:
\begin{numlstlisting}[name=SGrun]
  std::vector<Constraint<T>*> upward_constraints_1;
  upward_constraints_1.push_back(&low_scale_constraint_1);
  std::vector<Constraint<T>*> downward_constraints_1;
  downward_constraints_1.push_back(&susy_scale_constraint_1);
  downward_constraints_1.push_back(&low_scale_constraint_1);
\end{numlstlisting}
and initializes the \code{MSSMD5O} object:
\begin{numlstlisting}[name=SGrun,language=C++]
  model_1.clear();                 // of type MSSMD5O<T>
  model_1.set_input_parameters(input_1);
  model_1.do_calculate_sm_pole_masses(calculate_sm_masses);
\end{numlstlisting}
Likewise for \code{MSSMRHN}:
\begin{numlstlisting}[name=SGrun,language=C++]
  std::vector<Constraint<T>*> upward_constraints_2;
  upward_constraints_2.push_back(&high_scale_constraint_2);
  std::vector<Constraint<T>*> downward_constraints_2;
  downward_constraints_2.push_back(&high_scale_constraint_2);
  model_2.clear();                 // of type MSSMRHN<T>
  model_2.set_input_parameters(input_2);
\end{numlstlisting}
Note that \code{model_2} does not have to calculate the pole masses of
the SM particles since it is active only above $M_\nu$
which is assumed to be much higher than the weak scale.
To test the convergence of both models,
one may construct a composite convergence tester
out of auto-generated
\code{MSSMD5O_convergence_tester} and \code{MSSMRHN_convergence_tester}:
\begin{numlstlisting}[name=SGrun]
  MSSMD5O_convergence_tester<T> convergence_tester_1(&model_1, precision_goal);
  MSSMRHN_convergence_tester<T> convergence_tester_2(&model_2, precision_goal);
  if (max_iterations > 0) {
    convergence_tester_1.set_max_iterations(max_iterations);
    convergence_tester_2.set_max_iterations(max_iterations);
  }
  Composite_convergence_tester<T> convergence_tester;
  convergence_tester.add_convergence_tester(&convergence_tester_1);
  convergence_tester.add_convergence_tester(&convergence_tester_2);
\end{numlstlisting}
On construction,
the initial guesser accepts the following parameters
including the two model objects:
\begin{numlstlisting}[name=SGrun]
  MSSMD5O_MSSMRHN_initial_guesser<T> initial_guesser
    (&model_1, &model_2, input_1, oneset,
     low_scale_constraint_1, susy_scale_constraint_1, high_scale_constraint_2,
     matching);
\end{numlstlisting}
The code of the above class shall be presented later on.
One then passes
\code{convergence_tester} and \code{initial_guesser}
to \code{solver}, the \code{RGFlow} object,
along with the precision specification:
\begin{numlstlisting}[name=SGrun,language=C++]
  Two_scale_increasing_precision precision(10.0, precision_goal);
  solver.reset();                  // of type RGFlow<T>
  solver.set_convergence_tester(&convergence_tester);
  solver.set_running_precision(&precision);
  solver.set_initial_guesser(&initial_guesser);
\end{numlstlisting}
Finally,
one is ready to construct the tower of effective theories
by adding to \code{solver}
each model plus the associated list of constraints
optionally accompanied by a matching object:
\begin{numlstlisting}[name=SGrun]
  solver.add_model(&model_1, &matching, upward_constraints_1, downward_constraints_1);
  solver.add_model(&model_2, upward_constraints_2, downward_constraints_2);
\end{numlstlisting}
The order of addition is from the lowest scale to the highest.
Notice in line 49
that the matching object between \code{model_1} and
\code{model_2} is given when one adds the former, i.e.\ the lower-energy model.
It then remains to solve the boundary value problem:
\begin{numlstlisting}[name=SGrun]
  high_scale_2 = susy_scale_1 = low_scale_1 = 0; matching_scale = 0;
  solver.solve();
\end{numlstlisting}
After the solution is found,
one can obtain the resulting low-energy spectrum.
Since \code{model_1} is in contact with the lowest energy,
let it calculate the spectrum:
\begin{numlstlisting}[name=SGrun,language=C++]
  susy_scale_1 = susy_scale_constraint_1.get_scale();
  model_1.run_to(susy_scale_1);    // of type MSSMD5O<T>
  model_1.calculate_spectrum();
  if (!is_zero(parameter_output_scale_1))
    model_1.run_to(parameter_output_scale_1);
}
\end{numlstlisting}
In lines 56--57,
the scale is optionally brought to the value at which
one wishes to get the \DRbar\ parameters.

One needs to write the matching class for a particular pair of
models from scratch.
It shall be based on the abstract class \code{Matching<Two_scale>}
that comes with \fs.
In the present example, the class is declared in the header
\code{MSSMD5O_MSSMRHN_two_scale_matching.hpp}:
\begin{numlstlisting}
template<> class MSSMD5O_MSSMRHN_matching<Two_scale> : public Matching<Two_scale> {
public:
  MSSMD5O_MSSMRHN_matching();
  MSSMD5O_MSSMRHN_matching(const MSSMD5O_input_parameters&);
  void match_low_to_high_scale_model();
  void match_high_to_low_scale_model();
  double get_scale() const;
  void set_models(Two_scale_model *lower, Two_scale_model *upper);
  double get_initial_scale_guess() const;
  void set_lower_input_parameters(const MSSMD5O_input_parameters&);
  void set_scale(double);
  void reset();
private:
  MSSMD5O<Two_scale> *lower;
  MSSMRHN<Two_scale> *upper;
  void make_initial_scale_guess();
  void update_scale();
  ...
};
\end{numlstlisting}
As lines 4 and 10 indicate,
this class takes an \code{MSSMD5O_input_parameters} object as input.
The low-energy neutrino data therein is referenced by
\code{make_initial_scale_guess}
starting from line 23 of \code{MSSMD5O_MSSMRHN_two_scale_matching.cpp}:
\begin{numlstlisting}
void MSSMD5O_MSSMRHN_matching<Two_scale>::invert_seesaw_formula
(const Eigen::Matrix3d& WOp, const Eigen::Vector3d& YvDiag,
 Eigen::Matrix3d& Yv, Eigen::Matrix3d& Mv)
{
  Eigen::Matrix3cd uh;
  Eigen::Array3d s;
  fs_diagonalize_symmetric(WOp, s, uh);
  Eigen::Matrix3d U = uh.adjoint().real();
  Eigen::Vector3d YvDiagInv(1, 1, 1);
  YvDiagInv.array() /= YvDiag.array();
  Eigen::Matrix3d YvInv = U * YvDiagInv.asDiagonal();
  Mv = (YvInv.transpose() * WOp * YvInv).inverse();
  Yv = YvDiag.asDiagonal() * U.adjoint();
}

void MSSMD5O_MSSMRHN_matching<Two_scale>::set_lower_input_parameters
(const MSSMD5O_input_parameters& inputPars_)
{
  inputPars = inputPars_;
  make_initial_scale_guess();
}

void MSSMD5O_MSSMRHN_matching<Two_scale>::make_initial_scale_guess()
{
  Eigen::Matrix3d WOp;
  // fill WOp with elements in terms of data from inputPars;
  Eigen::Vector3d YvDiag;
  YvDiag << inputPars.YvDiag1, inputPars.YvDiag2, inputPars.YvDiag3;
  Eigen::Matrix3d Yv;
  Eigen::Matrix3d Mv;
  invert_seesaw_formula(WOp, YvDiag, Yv, Mv);
  double RHN_scale = pow(abs(Mv.determinant()), 1.0/3);
  scale = initial_scale_guess = RHN_scale;
}
\end{numlstlisting}
To guess the matching scale, this function estimates
the right-handed neutrino mass matrix \code{Mv}
from \code{WOp}, the dimension-5 operator coefficients,
and \code{YvDiag}, the neutrino Yukawa eigenvalues,
by calling \code{invert_seesaw_formula}.
Since the mapping,
$\mbox{\code{WOp}} \rightarrow (\mbox{\code{Yv}}, \mbox{\code{Mv}})$,
is not unique, \code{invert_seesaw_formula}
opts to impose the following additional constraints:
the eigenvalues of \code{Yv} are fixed to the user input (line 10),
and all mixing in \code{WOp} originates from the rotation of the SU(2)
doublets (line 11).

The actual matching process takes place in the two functions
\code{match_low_to_high_scale_model} and \code{match_high_to_low_scale_model}:
\begin{numlstlisting}
void MSSMD5O_MSSMRHN_matching<Two_scale>::match_low_to_high_scale_model()
{
  Eigen::Vector3d YvDiag;
  YvDiag << inputPars.YvDiag1, inputPars.YvDiag2, inputPars.YvDiag3;
  Eigen::Matrix3d Yv;
  Eigen::Matrix3d Mv;
  invert_seesaw_formula(lower->get_WOp(), YvDiag, Yv, Mv);
  upper->set_Yv(Yv);
  upper->set_Mv(Mv);

  upper->set_Yd(lower->get_Yd());
  // copy rest of couplings from lower to upper;
  upper->set_scale(lower->get_scale());
}

void MSSMD5O_MSSMRHN_matching<Two_scale>::match_high_to_low_scale_model()
{
  update_scale();

  const auto& Yv = upper->get_Yv();
  const auto& Mv = upper->get_Mv();
  lower->set_WOp(Yv.transpose() * Mv.inverse() * Yv);

  lower->set_Yd(upper->get_Yd());
  // copy rest of couplings from upper to lower;
  lower->set_scale(upper->get_scale());
}

void MSSMD5O_MSSMRHN_matching<Two_scale>::update_scale()
{
  double RHN_scale = pow(abs(upper->get_Mv().determinant()), 1.0/3);
  scale = RHN_scale;
}
\end{numlstlisting}
For the low-to-high matching, \code{invert_seesaw_formula} is called in line 7,
this time with \code{WOp} at the matching scale.
The high-to-low matching function contains the well-known
see-saw formula in line 22.
The matching scale is updated at each iteration
to be the geometric mean of the running \code{Mv} eigenvalues
in lines 31--32.

%
The last missing piece is the initial guesser.
One can extend the already available
\code{MSSMD5O_initial_guesser} class. 
The essential task is done by the following member function:
\begin{numlstlisting}[name=MSSMD5O_MSSMRHN_two_scale_initial_guesser.cpp]
void MSSMD5O_MSSMRHN_initial_guesser<Two_scale>::guess()
{
  // guess SUSY couplings in model-1 at low energy;

  const double low_scale_guess_1 = low_constraint_1.get_initial_scale_guess();
  const double high_scale_guess_2 = high_constraint_2.get_initial_scale_guess();
  const double matching_scale_guess = matching.get_initial_scale_guess();
\end{numlstlisting}
Compared to the MSSMD5O case, the differences are that the type of
\code{high_constraint_2} is \linebreak {\code{MSSMRHN_high_scale_constraint<Two_scale>} and that \code{matching_scale_guess} has been inserted.  Due to this intermediate scale, the initial run-up is divided into two steps, with a matching procedure in-between:
\begin{numlstlisting}[name=MSSMD5O_MSSMRHN_two_scale_initial_guesser.cpp,language=C++]
  model_1->run_to(matching_scale_guess); // of type MSSMD5O<Two_scale>
  matching.set_models(model_1, model_2);
  matching.match_low_to_high_scale_model();
  model_2->run_to(high_scale_guess_2);   // of type MSSMRHN<Two_scale>
\end{numlstlisting}
The high-scale constraints are applied to \code{model_2},
the higher-energy model,
and the remaining undetermined parameters are guessed:
\begin{numlstlisting}[name=MSSMD5O_MSSMRHN_two_scale_initial_guesser.cpp]
  high_constraint_2.set_model(model_2);
  high_constraint_2.apply();
  model_2->set_Mu(1.0); model_2->set_BMu(0.0);
\end{numlstlisting}
The initial two-step run-down again involves a matching process:
\begin{numlstlisting}[name=MSSMD5O_MSSMRHN_two_scale_initial_guesser.cpp]
  model_2->run_to(matching_scale_guess);
  matching.match_high_to_low_scale_model();
  model_1->run_to(low_scale_guess_1);
\end{numlstlisting}
At the low scale where \code{MSSMD5O} is valid,
the code is the same as in \code{MSSMD5O_initial_guesser}:
\begin{numlstlisting}[name=MSSMD5O_MSSMRHN_two_scale_initial_guesser.cpp]
  model_1->solve_ewsb_tree_level();
  model_1->calculate_DRbar_masses();
  model_1->set_thresholds(3); model_1->set_loops(2);
}
\end{numlstlisting}

Finally,
one prescribes the additional input parameters in the SLHA input file:
\begin{numlstlisting}
Block EXTPAR  # Input parameters
    1	5.0E-11		     # mv1
    2	5.07523E-11	     # mv2
    3	6.96419E-11	     # mv3
    4	0.586168	     # ThetaV12
    5	0.157512	     # ThetaV13
    6	0.705053	     # ThetaV23
    7	0.6		     # YvDiag1
    8	0.8		     # YvDiag2
    9	1.0		     # YvDiag3
Block BMvIN                  # right-handed sneutrino bilinear terms
1 1  1.000000E+02            # BMv(1,1)
...                          # remaining 8 entries
\end{numlstlisting}
The file in the package contains the values 
consistent with
the observed neutrino mass-squared differences and mixing angles
\cite{Beringer:1900zz}.

For further details, browse the directory \code{examples/tower/}.
One can build and run the example therein by:
\begin{numlstlisting}
$ cd examples/tower
$ make
$ ./run_tower.x --slha-input-file=LesHouches.in.tower
\end{numlstlisting}
In the output,
a part of the main interest is:
\begin{numlstlisting}
Block MSL2 Q= 8.82028104E+02
  1  1     1.26231305E+05   # ml2(1,1)
  1  2    -5.08365953E+02   # ml2(1,2)
  1  3     4.67463327E+01   # ml2(1,3)
  2  1    -5.08365953E+02   # ml2(2,1)
  2  2     1.25328663E+05   # ml2(2,2)
  2  3    -6.93488070E+02   # ml2(2,3)
  3  1     4.67463327E+01   # ml2(3,1)
  3  2    -6.93488070E+02   # ml2(3,2)
  3  3     1.24229816E+05   # ml2(3,3)
\end{numlstlisting}
This result demonstrates the well-known effect on
the off-diagonal slepton mass matrix elements
from a non-trivial flavour structure of \code{Yv}
\cite{Borzumati:1986qx}.
This leads in turn to the slepton mixing
matrices, \code{ZE} and \code{ZV},
which contain inter-generational mixings apart from the
generic left-right mixings.

\subsection{Integrating custom-built C++ components}
\label{sec:integrating-custom-built}

In \secref{sec:changing boundary conditions},
it was explained how one can let \fs generate
an alternative boundary condition class
by authoring a model file.
Nonetheless, the way to employ this class at the C++ level
might still remain obscure to the reader
since \fs automatically took care of it.
Here, an example shall be exhibited
with the emphasis on the modular C++ code structure
that helps such programming tasks.
Concretely,
the auto-generated low-energy boundary condition on the MSSM
shall be modified so that
$\alpha_{\text{s},\text{susy}}^{\text{\DRbar}}$
is determined from
$\alpha_{\text{s},\text{SM}}^{(5),\text{\MSbar}}$
by means of a two-loop matching.
This shall be accompanied by an improvement
of the $g_3$ $\beta$-function to the three-loop accuracy.

The first step is to alter
the model class which evaluates the $\beta$-functions.
Thanks to the \code{beta()} method being virtual,
one can override it conveniently by deriving a class from \code{MSSM}.
The declaration might look like:
\begin{numlstlisting}
#include "MSSM_two_scale_model.hpp"

template<>
class MSSMcbs<Two_scale> : public MSSM<Two_scale> {
public:
  explicit MSSMcbs(const MSSM_input_parameters& input_ = MSSM_input_parameters());
  virtual ~MSSMcbs();
  virtual Eigen::ArrayXd beta() const;
  MSSM_soft_parameters calc_beta() const;
};
\end{numlstlisting}
where the name \code{MSSMcbs} is an abbreviation of
the MSSM with custom-built $\beta$'s.
Note that the objects for \code{MSSM} are reused where possible:
\code{MSSM_input_parameters} in line 6 as well as
\code{MSSM_soft_parameters} in line 9.
This saves the programmer from excessive duplication of codes.
The member function definitions read:
\begin{numlstlisting}
Eigen::ArrayXd MSSMcbs<Two_scale>::beta() const
{
  return calc_beta().get();
}

MSSM_soft_parameters MSSMcbs<Two_scale>::calc_beta() const
{
  MSSM_soft_parameters betas(MSSM<Two_scale>::calc_beta());
  if (get_loops() <= 2) return betas;
  double bg33 = /* formula in terms of g1, g2, g3, Yu, Yd, Ye */;
  betas.set_g3(betas.get_g3() + Power(oneOver16PiSqr,3) * g3 * bg33);
  return betas;
}
\end{numlstlisting}
The full C++ expression of \code{bg33},
used in \code{MSSMcbs_two_scale_model.cpp},
has been adapted from the code by Jack and Jones \cite{3-loop MSSM betas}.
This already completes the amendment of the $g_3$ $\beta$-function.

The next step is to write a substitute for
the low-energy boundary condition class.
It must be declared as a descendant of \code{Constraint<Two_scale>}
whose function is described in \secref{sec:boundary-conditions}:
\begin{numlstlisting}
#include "two_scale_constraint.hpp"

template<>
class MSSMcbs_low_scale_constraint<Two_scale> : public Constraint<Two_scale> {
public:
  MSSMcbs_low_scale_constraint(const MSSM_input_parameters&, const QedQcd&);
  virtual ~MSSMcbs_low_scale_constraint();
  void set_threshold_corrections(unsigned);
  ...
private:
  MSSMcbs<Two_scale>* model;
  QedQcd oneset;
  double new_g3;
  unsigned threshold_corrections;
  void calculate_DRbar_gauge_couplings();
  double calculate_alS5DRbar_over_alS5MSbar(double) const;
  double calculate_zeta_g_QCD_2(double) const;
  double calculate_zeta_g_SUSY_2(double) const;
  ...
};
\end{numlstlisting}
In line 11,
the type of \code{model} has been adapted to the new model.
In fact, this class should work even if \code{model} remained
a pointer to \code{MSSM<Two_scale>} because of inheritance.
The main additions to \code{MSSM_low_scale_constraint}
in \code{models/MSSM/} are the member functions in lines 16--18,
which evaluate the two-loop matching coefficients
from Ref.~\cite{Harlander:2005wm}.
The following member function then performs
the two-step decoupling as reported in this reference:
\begin{numlstlisting}[language=C++]
void MSSMcbs_low_scale_constraint<Two_scale>::calculate_DRbar_gauge_couplings()
{
  ...
  double alpha_s = oneset.displayAlpha(ALPHAS);
  double alS5DRbar_over_alS5MSbar = 1;
  double zeta_g_QCD_2 = 1;
  double zeta_g_SUSY_2 = 1;
  if (model->get_thresholds()) {
    alS5DRbar_over_alS5MSbar = calculate_alS5DRbar_over_alS5MSbar(alpha_s);
    alpha_s *= alS5DRbar_over_alS5MSbar;  // alS5MSbar -> alS5DRbar
    zeta_g_QCD_2 = calculate_zeta_g_QCD_2(alpha_s);
    alpha_s /= zeta_g_QCD_2;		  // alS5DRbar -> alS6DRbar
    zeta_g_SUSY_2 = calculate_zeta_g_SUSY_2(alpha_s);
    alpha_s /= zeta_g_SUSY_2;		  // alS6DRbar -> alS6DRbarMSSM
    ...
  }
  new_g3 = Sqrt(4*Pi * alpha_s);
  ...
}
\end{numlstlisting}
Finally,
one can integrate the new boundary condition class
\code{MSSMcbs_low_scale_constraint} together with
the new model \code{MSSMcbs}
into the spectrum generator in a straightforward manner.
They should supersede
\code{MSSM_low_scale_constraint} and \code{MSSM}, respectively.
The replacement should be carried out in
those objects that depend on these classes,
i.e.\ the initial guesser:
\begin{numlstlisting}
template<>
class MSSMcbs_initial_guesser<Two_scale> : public Initial_guesser<Two_scale> {
public:
  MSSMcbs_initial_guesser(MSSMcbs<Two_scale>*,
                          const MSSM_input_parameters&,
                          const QedQcd&,
                          const MSSMcbs_low_scale_constraint<Two_scale>&,
                          const MSSM_susy_scale_constraint<Two_scale>&,
                          const MSSM_high_scale_constraint<Two_scale>&);
  ...
private:
  MSSMcbs<Two_scale>* model;
  MSSM_input_parameters input_pars;
  QedQcd oneset;
  MSSMcbs_low_scale_constraint<Two_scale> low_constraint;
  MSSM_susy_scale_constraint<Two_scale> susy_constraint;
  MSSM_high_scale_constraint<Two_scale> high_constraint;
  ...
};
\end{numlstlisting}
as well as the spectrum generator object:
\begin{numlstlisting}
template <class T>
void MSSMcbs_spectrum_generator<T>::run(const QedQcd& oneset,
					const MSSM_input_parameters& input)
{
  ...
  MSSMcbs_initial_guesser<T> initial_guesser
    (&model, input, oneset,
     low_scale_constraint, susy_scale_constraint, high_scale_constraint);
  ...
}
\end{numlstlisting}
One can find a working realization of this example
in \code{examples/customized-betas/}.

\section{Speed test SLHA input file}
\label{sec:speed-test-slha-template-file}
\begin{lstlisting}
Block MODSEL                 # Select model
    6    0                   # flavour violation
    1    1                   # mSUGRA
Block SMINPUTS               # Standard Model inputs
    1   1.279180000e+02      # alpha^(-1) SM MSbar(MZ)
    2   1.166390000e-05      # G_Fermi
    3   1.189000000e-01      # alpha_s(MZ) SM MSbar
    4   9.118760000e+01      # MZ(pole)
    5   4.200000000e+00      # mb(mb) SM MSbar
    6   1.709000000e+02      # mtop(pole)
    7   1.777000000e+00      # mtau(pole)
Block SOFTSUSY               # SOFTSUSY specific inputs
    1   1.000000000e-04      # tolerance
    2   2                    # up-quark mixing (=1) or down (=2)
    3   0                    # printout
    5   1                    # 2-loop running
    7   2                    # EWSB and Higgs mass loop order
Block FlexibleSUSY
    0   1.000000000e-04      # precision goal
    1   0                    # max. iterations (0 = automatic)
    2   0                    # algorithm (0 = two_scale, 1 = lattice)
    3   0                    # calculate SM pole masses
    4   2                    # pole mass loop order
    5   2                    # EWSB loop order
    6   2                    # beta-functions loop order
    7   1                    # threshold corrections loop order
    8   1                    # Higgs 2-loop corrections O(alpha_t alpha_s)
    9   1                    # Higgs 2-loop corrections O(alpha_b alpha_s)
   10   1                    # Higgs 2-loop corrections O(alpha_t^2 + alpha_t alpha_b + alpha_b^2)
   11   1                    # Higgs 2-loop corrections O(alpha_tau^2)
Block SPhenoInput            # SPheno specific input
    1  -1                    # error level
    2   1                    # SPA conventions
    11  0                    # calculate branching ratios
    13  0                    # include 3-Body decays
    12  1.000E-04            # write only branching ratios larger than this value
    31  -1                   # fixed GUT scale (-1: dynamical GUT scale)
    32  0                    # Strict unification
    34  1.000E-04            # Precision of mass calculation
    35  40                   # Maximal number of iterations
    37  1                    # Set Yukawa scheme
    38  2                    # 1- or 2-Loop RGEs
    50  1                    # Majorana phases: use only positive masses
    51  0                    # Write Output in CKM basis
    52  0                    # Write spectrum in case of tachyonic states
    55  1                    # Calculate one loop masses
    57  0                    # Calculate low energy constraints
    60  0                    # Include possible, kinetic mixing
    65  1                    # Solution tadpole equation
    75  0                    # Write WHIZARD files
    76  0                    # Write HiggsBounds file
    86  0.                   # Maximal width to be counted as invisible in Higgs decays
    510 0.                   # Write tree level values for tadpole solutions
    515 0                    # Write parameter values at GUT scale
    520 0.                   # Write effective Higgs couplings (HiggsBounds blocks)
    525 0.                   # Write loop contributions to diphoton decay of Higgs
Block MINPAR
    1   [50..1000]           # m0(MX)
    2   [50..1000]           # m12(MX)
    3   [1..100]             # tan(beta)(MZ) DRbar
    4   {-1,+1}              # sign(mu)
    5   [-1000..1000]        # A0(MX)
\end{lstlisting}


\clearpage

\begin{thebibliography}{100}
\bibitem{Coleman:1967ad} 
  S.~R.~Coleman and J.~Mandula,
  Phys.\ Rev.\  {\bf 159}, 1251 (1967).
\bibitem{Haag:1974qh} 
  R.~Haag, J.~T.~Lopuszanski and M.~Sohnius,
  Nucl.\ Phys.\ B {\bf 88}, 257 (1975).
\bibitem{Weinberg:1975gm} 
  S.~Weinberg,
  Phys.\ Rev.\ D {\bf 13}, 974 (1976).
\bibitem{Weinberg:1979bn}
  S.~Weinberg,
  Phys.\ Rev.\ D {\bf 19} (1979) 1277.
\bibitem{Gildener:1976ai} 
  E.~Gildener,
  Phys.\ Rev.\ D {\bf 14}, 1667 (1976).
\bibitem{Susskind:1978ms}
  L.~Susskind,
  Phys.\ Rev.\ D {\bf 20} (1979) 2619.
\bibitem{'tHooft:1980xb} 
  G.~'t Hooft, C.~Itzykson, A.~Jaffe, H.~Lehmann, P.~K.~Mitter, I.~M.~Singer and R.~Stora,
  NATO Adv.\ Study Inst.\ Ser.\ B Phys.\  {\bf 59}, pp.1 (1980).

\bibitem{Langacker:1990jh} 
  P.~Langacker,
  In *Boston 1990, Proceedings, Particles, strings and cosmology* 237-269 and Pennsylvania Univ. Philadelphia - UPR-0435T (90,rec.Oct.) 33 p. (015721) (see HIGH ENERGY PHYSICS INDEX 29 (1991) No. 9950)

\bibitem{Ellis:1990wk} 
  J.~R.~Ellis, S.~Kelley and D.~V.~Nanopoulos,
  Phys.\ Lett.\ B {\bf 260}, 131 (1991).
\bibitem{Amaldi:1991cn} 
  U.~Amaldi, W.~de Boer and H.~Furstenau,
  Phys.\ Lett.\ B {\bf 260}, 447 (1991).
\bibitem{Langacker:1991an} 
  P.~Langacker and M.~-x.~Luo,
  Phys.\ Rev.\ D {\bf 44}, 817 (1991).
\bibitem{Giunti:1991ta} 
  C.~Giunti, C.~W.~Kim and U.~W.~Lee,
  Mod.\ Phys.\ Lett.\ A {\bf 6}, 1745 (1991).


\bibitem{Goldberg:1983nd} 
  H.~Goldberg,
  Phys.\ Rev.\ Lett.\  {\bf 50}, 1419 (1983)
  [Erratum-ibid.\  {\bf 103}, 099905 (2009)].
\bibitem{Ellis:1983ew} 
  J.~R.~Ellis, J.~S.~Hagelin, D.~V.~Nanopoulos, K.~A.~Olive and M.~Srednicki,
  Nucl.\ Phys.\ B {\bf 238}, 453 (1984).

\bibitem{Girardello:1981wz}
  L.~Girardello and M.~T.~Grisaru,
  Nucl.\ Phys.\ B {\bf 194} (1982) 65.
\bibitem{Chung:2003fi} D.~J.~H.~Chung, L.~L.~Everett, G.~L.~Kane,
  S.~F.~King, J.~D.~Lykken and L.~T.~Wang,
  Phys.\ Rept.\  {\bf 407}, 1 (2005)
  [hep-ph/0312378].

\bibitem{Allanach:2001kg} 
  B.~C.~Allanach,
  Comput.\ Phys.\ Commun.\  {\bf 143}, 305 (2002)
  [hep-ph/0104145].
\bibitem{Porod:2003um} 
  W.~Porod,
  Comput.\ Phys.\ Commun.\  {\bf 153}, 275 (2003)
  [hep-ph/0301101].

\bibitem{Djouadi:2002ze} 
  A.~Djouadi, J.~-L.~Kneur and G.~Moultaka,
  Comput.\ Phys.\ Commun.\  {\bf 176}, 426 (2007)
  [hep-ph/0211331].

\bibitem{Baer:1993ae} 
  H.~Baer, F.~E.~Paige, S.~D.~Protopopescu and X.~Tata,
  hep-ph/9305342.

\bibitem{Chowdhury:2011zr}
  D.~Chowdhury, R.~Garani and S.~K.~Vempati,
  Comput.\ Phys.\ Commun.\  {\bf 184} (2013) 899
  [arXiv:1109.3551 [hep-ph]].

\bibitem{Ellwanger:2004xm} 
  U.~Ellwanger, J.~F.~Gunion and C.~Hugonie,
  JHEP {\bf 0502}, 066 (2005)
  [hep-ph/0406215].

\bibitem{Ellwanger:2005dv} 
  U.~Ellwanger and C.~Hugonie,
  Comput.\ Phys.\ Commun.\  {\bf 175}, 290 (2006)
  [hep-ph/0508022].


\bibitem{Ellwanger:2006rn} 
  U.~Ellwanger and C.~Hugonie,
  Comput.\ Phys.\ Commun.\  {\bf 177}, 399 (2007)
  [hep-ph/0612134].

\bibitem{Ellwanger:2008py} 
  U.~Ellwanger, C.-C.~Jean-Louis and A.~M.~Teixeira,
  JHEP {\bf 0805}, 044 (2008)
  [arXiv:0803.2962 [hep-ph]].




\bibitem{Allanach:2013kza} 
  B.~C.~Allanach, P.~Athron, L.~C.~Tunstall, A.~Voigt and A.~G.~Williams,
  Comput.\ Phys.\ Commun.\  {\bf 185}, 2322 (2014)
  [arXiv:1311.7659 [hep-ph]].
\bibitem{Ellwanger:2009dp} 
  U.~Ellwanger, C.~Hugonie and A.~M.~Teixeira,
  Phys.\ Rept.\  {\bf 496}, 1 (2010)
  [arXiv:0910.1785 [hep-ph]].
\bibitem{Maniatis:2009re} 
  M.~Maniatis,
  Int.\ J.\ Mod.\ Phys.\ A {\bf 25}, 3505 (2010)
  [arXiv:0906.0777 [hep-ph]].


\bibitem{Kim:1983dt} 
  J.~E.~Kim and H.~P.~Nilles,
  Phys.\ Lett.\ B {\bf 138}, 150 (1984).
\bibitem{King:2014nza} 
  S.~F.~King, A.~Merle, S.~Morisi, Y.~Shimizu and M.~Tanimoto,
  arXiv:1402.4271 [hep-ph].
\bibitem{King:2008qb} 
  S.~F.~King, R.~Luo, D.~J.~Miller and R.~Nevzorov,
  JHEP {\bf 0812}, 042 (2008)
  [arXiv:0806.0330 [hep-ph]].
\bibitem{Aad:2013wta} 
  G.~Aad {\it et al.}  [ATLAS Collaboration],
  JHEP {\bf 1310}, 130 (2013)
  [arXiv:1308.1841 [hep-ex]].
\bibitem{Chatrchyan:2014lfa}
  S.~Chatrchyan {\it et al.}  [CMS Collaboration],
  arXiv:1402.4770 [hep-ex].
\bibitem{ATLAS:2012ae} 
  G.~Aad {\it et al.}  [ATLAS Collaboration],
  Phys.\ Lett.\ B {\bf 710}, 49 (2012)
  [arXiv:1202.1408 [hep-ex]].
\bibitem{Chatrchyan:2012tx} 
  S.~Chatrchyan {\it et al.}  [CMS Collaboration],
  Phys.\ Lett.\ B {\bf 710}, 26 (2012)
  [arXiv:1202.1488 [hep-ex]].

\bibitem{Fayet:1977yc}
  P.~Fayet,
  Phys.\ Lett.\ B {\bf 69} (1977) 489.

\bibitem{Cvetic:1997ky}
  M.~Cvetic, D.~A.~Demir, J.~R.~Espinosa, L.~L.~Everett and P.~Langacker,
  Phys.\ Rev.\ D {\bf 56} (1997) 2861
   [Erratum-ibid.\ D {\bf 58} (1998) 119905]
  [hep-ph/9703317].

\bibitem{Langacker:2008yv} 
  P.~Langacker,
  Rev.\ Mod.\ Phys.\  {\bf 81}, 1199 (2009)
  [arXiv:0801.1345 [hep-ph]].



\bibitem{King:2005jy} 
  S.~F.~King, S.~Moretti and R.~Nevzorov,
  Phys.\ Rev.\ D {\bf 73}, 035009 (2006)
  [hep-ph/0510419].

\bibitem{Athron:2010zz} 
  P.~Athron, J.~P.~Hall, R.~Howl, S.~F.~King, D.~J.~Miller, S.~Moretti and R.~Nevzorov,
  Nucl.\ Phys.\ Proc.\ Suppl.\  {\bf 200-202}, 120 (2010).

\bibitem{Nevzorov:2012hs} 
  R.~Nevzorov,
  Phys.\ Rev.\ D {\bf 87}, 015029 (2013)
  [arXiv:1205.5967 [hep-ph]].



\bibitem{Batra:2003nj} 
  P.~Batra, A.~Delgado, D.~E.~Kaplan and T.~M.~P.~Tait,
  JHEP {\bf 0402}, 043 (2004)
  [hep-ph/0309149].

\bibitem{Bharucha:2013ela} 
  A.~Bharucha, A.~Goudelis and M.~McGarrie,
  arXiv:1310.4500 [hep-ph].

\bibitem{Staub:2010ty} 
  F.~Staub, W.~Porod and B.~Herrmann,
  JHEP {\bf 1010}, 040 (2010)
  [arXiv:1007.4049 [hep-ph]].

\bibitem{Staub:2009bi} 
  F.~Staub,
  Comput.\ Phys.\ Commun.\  {\bf 181}, 1077 (2010)
  [arXiv:0909.2863 [hep-ph]].

\bibitem{Staub:2010jh} 
  F.~Staub,
  Comput.\ Phys.\ Commun.\  {\bf 182}, 808 (2011)
  [arXiv:1002.0840 [hep-ph]].

\bibitem{Staub:2012pb} 
  F.~Staub,
  Computer Physics Communications {\bf 184}, pp. 1792 (2013)
  [Comput.\ Phys.\ Commun.\  {\bf 184}, 1792 (2013)]
  [arXiv:1207.0906 [hep-ph]].

\bibitem{Staub:2013tta} 
  F.~Staub,
  arXiv:1309.7223 [hep-ph].

\bibitem{Degrassi:2001yf}
  G.~Degrassi, P.~Slavich and F.~Zwirner,
  Nucl.\ Phys.\ B {\bf 611} (2001) 403
  [hep-ph/0105096].

\bibitem{Brignole:2001jy}
  A.~Brignole, G.~Degrassi, P.~Slavich and F.~Zwirner,
  Nucl.\ Phys.\ B {\bf 631} (2002) 195
  [hep-ph/0112177].

\bibitem{Dedes:2002dy}
  A.~Dedes and P.~Slavich,
  Nucl.\ Phys.\ B {\bf 657} (2003) 333
  [hep-ph/0212132].

\bibitem{Brignole:2002bz}
  A.~Brignole, G.~Degrassi, P.~Slavich and F.~Zwirner,
  Nucl.\ Phys.\ B {\bf 643} (2002) 79
  [hep-ph/0206101].

\bibitem{Dedes:2003km}
  A.~Dedes, G.~Degrassi and P.~Slavich,
  Nucl.\ Phys.\ B {\bf 672} (2003) 144
  [hep-ph/0305127].

\bibitem{slhaea} F.S.~Thomas http://fthomas.github.io/slhaea/

\bibitem{hdecay}  A. Djouadi, M. Spira and P.M. Zerwas, Phys. Lett.  B {\bf 264}
(1991) 440 and  Z. Phys. C {\bf 70} (1996) 427;  M. Spira {\it et al.}, Nucl.\
Phys.\ B {\bf 453} (1995) 17;  A.~Djouadi, J.~Kalinowski and M.~Spira, Comput.
Phys. Commun. {\bf 108} (1998) 56; J.~M.~Butterworth, A.~Arbey, L.~Basso, S.~Belov, A.~Bharucha, F.~Braam, A.~Buckley and M.~Campanelli {\it et al.},
  arXiv:1003.1643 [hep-ph].
\bibitem{sdecay} 
  M.~Muhlleitner, A.~Djouadi and Y.~Mambrini,
  Comput.\ Phys.\ Commun.\  {\bf 168}, 46 (2005)
  [hep-ph/0311167].


\bibitem{susyhit} 
  A.~Djouadi, M.~M.~Muhlleitner and M.~Spira,
  Acta Phys.\ Polon.\ B {\bf 38}, 635 (2007)
  [hep-ph/0609292].

\bibitem{Wells:2003tf} 
  J.~D.~Wells,
  hep-ph/0306127.

\bibitem{ArkaniHamed:2004fb} 
  N.~Arkani-Hamed and S.~Dimopoulos,
  JHEP {\bf 0506}, 073 (2005)
  [hep-th/0405159].
\bibitem{Giudice:2004tc}
  G.~F.~Giudice and A.~Romanino,
  Nucl.\ Phys.\ B {\bf 699} (2004) 65
   [Erratum-ibid.\ B {\bf 706} (2005) 65]
  [hep-ph/0406088].

\bibitem{Degrassi:2009yq} 
  G.~Degrassi and P.~Slavich,
  Nucl.\ Phys.\ B {\bf 825}, 119 (2010)
  [arXiv:0907.4682 [hep-ph]].

\bibitem{eigen}
  Eigen library, version 3.1 \url{http://eigen.tuxfamily.org}.

\bibitem{Skands:2003cj}
  P.~Z.~Skands, B.~C.~Allanach, H.~Baer, C.~Balazs, G.~Belanger, F.~Boudjema, A.~Djouadi and R.~Godbole {\it et al.},
  JHEP {\bf 0407} (2004) 036
  [hep-ph/0311123].

\bibitem{Allanach:2008qq} 
  B.~C.~Allanach, C.~Balazs, G.~Belanger, M.~Bernhardt, F.~Boudjema, D.~Choudhury, K.~Desch and U.~Ellwanger {\it et al.},
  Comput.\ Phys.\ Commun.\  {\bf 180}, 8 (2009)
  [arXiv:0801.0045 [hep-ph]].



\bibitem{Athron:2007en}
  P.~Athron, S.~F.~King, D.~J.~Miller, S.~Moretti and R.~Nevzorov,
  J.\ Phys.\ Conf.\ Ser.\  {\bf 110} (2008) 072001
  [arXiv:0708.3248 [hep-ph]].



\bibitem{Kribs:2007ac}
  G.~D.~Kribs, E.~Poppitz and N.~Weiner,
  Phys.\ Rev.\ D {\bf 78} (2008) 055010
  [arXiv:0712.2039 [hep-ph]].

\bibitem{Staub:2008uz}
  F.~Staub,
  arXiv:0806.0538 [hep-ph].

\bibitem{Jones:1974pg}
  D.~R.~T.~Jones,
  Nucl.\ Phys.\ B {\bf 87} (1975) 127.

\bibitem{Jones:1983vk}
  D.~R.~T.~Jones and L.~Mezincescu,
  Phys.\ Lett.\ B {\bf 136} (1984) 242.

\bibitem{West:1984dg}
  P.~C.~West,
  Phys.\ Lett.\ B {\bf 137} (1984) 371.

\bibitem{Martin:1993yx}
  S.~P.~Martin and M.~T.~Vaughn,
  Phys.\ Lett.\ B {\bf 318} (1993) 331
  [hep-ph/9308222].

\bibitem{Yamada:1993ga}
  Y.~Yamada,
  Phys.\ Rev.\ Lett.\  {\bf 72} (1994) 25
  [hep-ph/9308304].

\bibitem{MV94} 
  S.~P.~Martin and M.~T.~Vaughn,
  Phys.\ Rev.\ D {\bf 50}, 2282 (1994)
  [Erratum-ibid.\ D {\bf 78}, 039903 (2008)]
  [hep-ph/9311340].



\bibitem{Yam94} 
  Y.~Yamada,
  Phys.\ Rev.\ D {\bf 50}, 3537 (1994)
  [hep-ph/9401241].
\bibitem{Jack:1994kd} 
  I.~Jack and D.~R.~T.~Jones,
  Phys.\ Lett.\ B {\bf 333}, 372 (1994)
  [hep-ph/9405233].

\bibitem{Jack:1994rk} 
  I.~Jack, D.~R.~T.~Jones, S.~P.~Martin, M.~T.~Vaughn and Y.~Yamada,
  Phys.\ Rev.\ D {\bf 50}, 5481 (1994)
  [hep-ph/9407291].

\bibitem{Fonseca:2011vn}
  R.~M.~Fonseca, M.~Malinsky, W.~Porod and F.~Staub,
  Nucl.\ Phys.\ B {\bf 854} (2012) 28
  [arXiv:1107.2670 [hep-ph]].

\bibitem{Goodsell:2012fm} 
  M.~D.~Goodsell,
  JHEP {\bf 1301}, 066 (2013)
  [arXiv:1206.6697 [hep-ph]].

\bibitem{Sperling:2013eva}
  M.~Sperling, D.~Stöckinger and A.~Voigt,
  JHEP {\bf 1307} (2013) 132
  [arXiv:1305.1548 [hep-ph]].

\bibitem{Sperling:2013xqa}
  M.~Sperling, D.~Stöckinger and A.~Voigt,
  JHEP {\bf 1401} (2014) 068
  [arXiv:1310.7629 [hep-ph]].




\bibitem{Beringer:1900zz}
  J.~Beringer {\it et al.}  [Particle Data Group Collaboration],
  Phys.\ Rev.\ D {\bf 86} (2012) 010001.

\bibitem{Hall:1980kf}
  L.~J.~Hall,
  Nucl.\ Phys.\ B {\bf 178} (1981) 75.

\bibitem{Avdeev:1997sz}
  L.~V.~Avdeev and M.~Y.~Kalmykov,
  Nucl.\ Phys.\ B {\bf 502} (1997) 419
  [hep-ph/9701308].

\bibitem{Bednyakov:2002sf}
  A.~Bednyakov, A.~Onishchenko, V.~Velizhanin and O.~Veretin,
  Eur.\ Phys.\ J.\ C {\bf 29} (2003) 87
  [hep-ph/0210258].

\bibitem{Baer:2002ek}
  H.~Baer, J.~Ferrandis, K.~Melnikov and X.~Tata,
  Phys.\ Rev.\ D {\bf 66} (2002) 074007
  [hep-ph/0207126].

\bibitem{Barger:1993gh} 
  V.~D.~Barger, M.~S.~Berger and P.~Ohmann,
  Phys.\ Rev.\ D {\bf 49}, 4908 (1994)
  [hep-ph/9311269].

\bibitem{NUHM}
  V.~Berezinsky, A.~Bottino, J.~R.~Ellis, N.~Fornengo, G.~Mignola and S.~Scopel,
  Astropart.\ Phys.\  {\bf 5} (1996) 1
  [hep-ph/9508249];
  P.~Nath and R.~L.~Arnowitt,
  Phys.\ Rev.\ D {\bf 56} (1997) 2820
  [hep-ph/9701301];
  A.~Bottino, F.~Donato, N.~Fornengo and S.~Scopel,
  Phys.\ Rev.\ D {\bf 63} (2001) 125003
  [hep-ph/0010203];
  V.~Bertin, E.~Nezri and J.~Orloff,
  JHEP {\bf 0302} (2003) 046
  [hep-ph/0210034];
%
  M.~Drees, M.~M.~Nojiri, D.~P.~Roy and Y.~Yamada,
  Phys.\ Rev.\ D {\bf 56} (1997) 276
   [Erratum-ibid.\ D {\bf 64} (2001) 039901]
  [hep-ph/9701219];
  M.~Drees, Y.~G.~Kim, M.~M.~Nojiri, D.~Toya, K.~Hasuko and T.~Kobayashi,
  Phys.\ Rev.\ D {\bf 63} (2001) 035008
  [hep-ph/0007202];
%
  J.~R.~Ellis, T.~Falk, G.~Ganis, K.~A.~Olive and M.~Schmitt,
  Phys.\ Rev.\ D {\bf 58} (1998) 095002
  [hep-ph/9801445];
  J.~R.~Ellis, T.~Falk, G.~Ganis and K.~A.~Olive,
  Phys.\ Rev.\ D {\bf 62} (2000) 075010
  [hep-ph/0004169];
%
  J.~R.~Ellis, K.~A.~Olive and Y.~Santoso,
  Phys.\ Lett.\ B {\bf 539} (2002) 107
  [hep-ph/0204192];
  J.~R.~Ellis, T.~Falk, K.~A.~Olive and Y.~Santoso,
  Nucl.\ Phys.\ B {\bf 652} (2003) 259
  [hep-ph/0210205].

\bibitem{see-saw}
  P.~Minkowski, \ptitle{
  $\mu \rightarrow e \gamma$ at a Rate of One Out of 1-Billion Muon Decays?,}
  Phys.\ Lett.\ B {\bf 67} (1977) 421;
  T.~Yanagida, \ptitle{
  Horizontal Symmetry And Masses Of Neutrinos,}
  Proc.\ of the
  Workshop on Unified Theories and the Baryon Number of the Universe,
  edited by O.~Sawada and A.~Sugamoto, KEK, Japan (1979) 95;
  M.~Gell-Mann, P.~Ramond and R.~Slansky, \ptitle{
  Complex Spinors and Unified Theories,}
  Supergravity, edited by F.~Nieuwenhuizen and
  D.~Friedman, North Holland, Amsterdam (1979) 315
  [arXiv:1306.4669 [hep-th]];
  R.~N.~Mohapatra and G.~Senjanovic, \ptitle{
  Neutrino Mass and Spontaneous Parity Nonconservation,}
  Phys.\ Rev.\ Lett.\  {\bf 44} (1980) 912.

\bibitem{Borzumati:1986qx}
  F.~Borzumati and A.~Masiero, \ptitle{
Large Muon- and electron-Number Nonconservation in Supergravity Theories,}
  Phys.\ Rev.\ Lett.\  {\bf 57} (1986) 961.

\bibitem{Allanach:2014nba}
  B.~C.~Allanach, A.~Bednyakov and R.~Ruiz de Austri,
  arXiv:1407.6130 [hep-ph].

\bibitem{Athron:2009ue} 
  P.~Athron, S.~F.~King, D.~J.~Miller, S.~Moretti and R.~Nevzorov,
  Phys.\ Lett.\ B {\bf 681}, 448 (2009)
  [arXiv:0901.1192 [hep-ph]].


\bibitem{Athron:2009bs}
  P.~Athron, S.~F.~King, D.~J.~Miller, S.~Moretti and R.~Nevzorov,
  Phys.\ Rev.\ D {\bf 80} (2009) 035009
  [arXiv:0904.2169 [hep-ph]].



\bibitem{Athron:2012pw}
  P.~Athron, D.~Stockinger and A.~Voigt,
  Phys.\ Rev.\ D {\bf 86} (2012) 095012
  [arXiv:1209.1470 [hep-ph]].


\bibitem{3-loop MSSM betas}
  I.~Jack and D.~R.~T.~Jones, \ptitle{
    3-loop MSSM beta functions,}
  \url{http://www.liv.ac.uk/~dij/betas}.

\bibitem{Harlander:2005wm}
  R.~Harlander, L.~Mihaila and M.~Steinhauser,
  Phys.\ Rev.\ D {\bf 72} (2005) 095009
  [hep-ph/0509048].

\end{thebibliography}
\end{document}